\documentclass[12pt]{article}

\pdfoutput=1

\usepackage[english]{babel}
\usepackage{fancyhdr}
\usepackage{amsmath}
\usepackage{amssymb}
\usepackage{amsfonts}
\usepackage{psfrag}
\usepackage[applemac]{inputenc}
\usepackage{graphicx,wrapfig}
\usepackage{pstricks}

\usepackage{slashed}            
\usepackage{subfig}             
\usepackage{xspace}				
\usepackage[sort&compress,square,comma,numbers]{natbib}
\usepackage{float}
\usepackage{dcolumn}
\usepackage{bm}
\usepackage{epstopdf}
\usepackage[
	colorlinks=true,
	citecolor=black,
	linkcolor=black,
	urlcolor=black,
	hypertexnames=false]{hyperref}
		
\newcommand{\arXiv}[2]{\href{http://arxiv.org/pdf/#1}{{\tt #2/#1}}}
\newcommand{\arXivold}[1]{\href{http://arxiv.org/pdf/#1}{{\tt #1}}}
\DeclareFontFamily{OT1}{pzc}{}
\DeclareFontShape{OT1}{pzc}{m}{it}{<-> s * [1.100] pzcmi7t}{}
\DeclareMathAlphabet{\mathpzc}{OT1}{pzc}{m}{it}
\newcommand{\QH}{\mathcal{H}}
\newcommand{\Qh}{\mathpzc{h}}

\newcommand{\beq}{\begin{eqnarray}}
\newcommand{\eeq}{\end{eqnarray}}
\newcommand{\bea}{\begin{eqnarray}}
\newcommand{\eea}{\end{eqnarray}}
\newcommand{\bag}{\begin{align}}
\newcommand{\eag}{\end{align}}

\newcommand{\ie}{$\textnormal{i.e.}$ }

\newcommand{\GeV}{\,\mathrm{GeV}}
\newcommand{\TeV}{\,\mathrm{TeV}}

\newcommand{\eq}[1]{Eq.~(\ref{#1})}

\newcommand{\vev}[1]{\langle {#1} \rangle}

\newcommand{\Lag}{\mathcal{L}}

\begin{document}

\baselineskip=18pt

\setcounter{footnote}{0}
\setcounter{figure}{0}
\setcounter{table}{0}

\begin{titlepage}

\hfill{CERN-PH-TH-2015-275}

\hfill{Saclay-t15/206}

\vspace{1cm}

\begin{center}
  \begin{LARGE}
    \begin{bf}
The Quantum Critical Higgs 
   \end{bf}
  \end{LARGE}
\end{center}
\vspace{0.1cm}
\begin{center}
\begin{large}
{\bf Brando Bellazzini$^{a,b}$, Csaba Cs\'aki$^c$, Jay Hubisz$^d$, Seung J. Lee$^{e,f}$,\\ Javi Serra$^g$, John Terning$^h$ \\}
\end{large}
  \vspace{0.5cm}
  \begin{it}

\begin{small}
$^{(a)}$Institut de Physique Th\'eorique, Universit\'e Paris Saclay, CEA, CNRS, F-91191 Gif-sur-Yvette, France
\vspace{0.2cm}\\
$^{(b)}$Dipartimento di Fisica e Astronomia, Universit\'a di Padova, 
Via Marzolo 8, I-35131 Padova, Italy
\vspace{0.2cm}\\
$^{(c)}$Department of Physics, LEPP, Cornell University, Ithaca, NY 14853, USA
\vspace{0.2cm}\\
$^{(d)}$Department of Physics, Syracuse University, Syracuse, NY 13244, USA
\vspace{0.2cm}\\
$^{(e)}$Department of Physics, Korea University, Seoul 136-713, Korea
\vspace{0.2cm}\\
$^{(f)}$School of Physics, Korea Institute for Advanced Study, Seoul 130-722, Korea
\vspace{0.2cm}\\
$^{(g)}$CERN, Theory Division, Geneva, Switzerland
\vspace{0.2cm}\\
$^{(h)}$Department of Physics, University of California, Davis, CA 95616
 \vspace{0.1cm}

\end{small}

\end{it}
\vspace{.5cm}
\end{center}

\begin{abstract} 
\medskip
\noindent

The appearance of the light Higgs boson at the LHC is difficult to explain, particularly in light of naturalness arguments in quantum field theory.
However light scalars can appear in condensed matter systems when parameters (like the amount of doping) are tuned to a critical point. At zero temperature these quantum critical points are directly analogous to the finely tuned standard model.  In this paper we explore a class of models with a Higgs near a quantum critical point that exhibits non-mean-field behavior. We discuss the parametrization of the effects of a Higgs emerging from such a critical point in terms of form factors, and present two simple realistic scenarios based on either generalized free fields or a 5D dual in AdS space.  For both of these models we consider the processes $gg\to ZZ$ and $gg\to hh$, which can be used to gain information about the Higgs scaling dimension and IR transition scale from the experimental data. 
 
\end{abstract}

\bigskip
\end{titlepage}

\section{Introduction} \label{intro}

The Higgs boson mass has been measured to be around 125 GeV by the LHC experiments. The appearance of a light scalar degree of freedom is quite unusual both in particle physics and in condensed matter systems~\footnote{This is true with the exception of (pseudo-)Nambu-Goldstone bosons (pNGBs), which arise as (nearly) massless scalar modes after the spontaneous breaking of a global symmetry of the Lagrangian. On the contrary, the light scalar we are discussing here is the excitation of the magnitude of the associated condensate rather than its phase.}. While there is no previous particle physics precedent, some condensed matter systems can produce a light scalar by tuning parameters 
close to a critical value where a continuous (second order) phase transition occurs. As the critical point is approached the correlation length diverges, which is an indication that the mass of the corresponding excitation approaches zero. At the critical point the system has an approximate scale invariance and at low energies we will see the universal behavior of some fixed point that constitutes the low-energy effective theory. If the system approaches a trivial fixed point then we find ``mean-field" critical exponents associated with the Landau-Ginzburg effective theory, and a light scalar excitation. The Higgs sector of the standard model (SM) is, in fact, precisely analogous to a Landau-Ginzburg theory. However, if the system is in the domain of attraction of a non-trivial fixed point then we find non-trivial critical exponents, and potentially no simple particle description. 

Phase transitions that occur at zero temperature as some other parameter is varied are referred to as quantum phase transitions (QPT's) since quantum fluctuations dominate over the more usual thermal fluctuations (see e.g.~\cite{Sachdev} and reference therein), and this is the case of interest for particle physics. Experimentally we know that the Higgs is much lighter than our theoretical expectations. In the SM varying the Higgs mass parameter in the Lagrangian provides for a continuous phase transition where the physical Higgs mass (and VEV) goes through zero, therefore in the SM we are extremely close to the quantum critical point \cite{qcpt} of a QPT with mean-field behavior. Indeed, if the SM is correct up to the Planck scale, then a change in 1 part in $10^{28}$ can push us through the phase transition, so we are very, very close to the critical point. If there is new physics beyond the SM then the relevant questions are: ``Does the underlying theory also have a QPT?" and ``If so, is it more interesting than mean-field theory?"

At a QPT the approximate scale invariant theory is characterized by the scaling dimensions, $\Delta$, of the gauge invariant operators. In the SM we have only small, perturbative corrections to the dimensions of Higgs-operators: $\Delta=1+ {\mathcal O}(\alpha/4 \pi)$, corresponding to mean-field behavior. The purpose of this paper is to present a general class of theories describing a Higgs field near a non-mean-field QPT, and explore the observable consequences. In such theories, in addition to the pole corresponding to the recently discovered Higgs boson, there can also be a Higgs continuum, which could potentially start not too far above the Higgs mass. The continuum represents additional states associated with the dynamics underlying the QPT, which we assume is described by a strongly coupled conformal field theory (CFT)~\footnote{The idea of using non-trivial fixed points and strongly coupled CFT's to understand the hierarchy problem goes back to variants of technicolor \cite{HoldomCFT,walking} and extends to composite models in AdS \cite{Agashe:2004rs} and beyond.}. The Higgs field can create all of these states, both the pole and the continuum. The pole itself could be just an elementary scalar that mixes with some states from the CFT. In this case the hierarchy problem will be just like in the SM. Another interesting possibility would be if the pole itself was a composite bound state of the CFT similarly to composite Higgs models~\cite{PGBHiggs}. Most of the discussion here will be general and will not make a distinction between these cases. 

One result of the presence of the continuum will be the appearance of form factors in couplings of the Higgs to the SM particles. Furthermore, associated with the dynamics of the non-trivial fixed point there will generically be extra states, which will decouple from low energies at some cutoff scale. Depending on how close these states are to the electroweak scale, their effects on the effective theory will be more or less important. This is just as in the SM supplemented by higher dimensional operators, with the extra information that the states of the QPT are expected to couple strongly to the Higgs field and, because of the Higgs' large anomalous dimension, a generic operator with a given number of Higgs insertions will be more irrelevant than the analog one in the weakly coupled case.

The phenomenology of these models will share some features with scenarios where the Higgs is involved with a conformal sector \cite{Unhiggs,Englert:2012dq,Heckman:2011bb,mixing,other}, however here we will try to formulate a general low-energy effective theory consistent with a QPT and no new massless particles.

As in the SM, our effective theory will not allow us to address the question of how Nature ends up tuned close to the QPT critical point. However, since we seem to be near such a critical point, it is worth considering  what effective theories can accommodate a light Higgs and still offer phenomenology distinct from the usual perturbative Higgs models. This is the case of a quantum critical Higgs (QCH) that we will explore in this paper.

The paper is organized as follows: we first discuss the general phenomenology of Higgs form factors, we then present two general classes of models for a Higgs near a QPT, and finally we discuss observable signals at the LHC that can allow us to extract the Higgs scaling dimension and IR transition scale.

\section{Form Factors for Higgs Phenomenology}

Our ultimate goal is to investigate scenarios where the Higgs is partially embedded into a strongly coupled sector.  We envision that such a sector is approximately conformal at scales well above the weak scale, as expected at a quantum critical point.  This would allow the type of scenario outlined in the Introduction:  the Higgs has a significant anomalous dimension and a continuum contribution to $n$-point functions. In order to understand what types of new physics effects could appear, we will first present a model-independent parametrization, based on form factors, of the various amplitudes controlling the main Higgs production and decay processes at the LHC.  
We will assume that the SM fermions, the massless gauge bosons, and the transverse parts of the $W$ and $Z$ are external to the CFT, or in other words, they are elementary states, while the Higgs (along with the Goldstone bosons associated with the longitudinal components of the $W$ and $Z$) originates from or is mixed with the strong sector, corresponding to a theory with spontaneously or explicitly broken conformal symmetry.  Note that in the case of a fully spontaneous breaking of conformal symmetry an additional massless scalar called the dilaton emerges as the Goldstone-boson for broken scale invariance. The idea that the Higgs pole at 125 GeV itself could be a dilaton has been entertained previously in~\cite{Witek, ourdilaton, Chacko, Kitano}. However, for realistic non-supersymmetric examples one also needs an explicit breaking of scale invariance to stabilize the symmetry breaking minimum, which generically pushes the dilaton mass to high values~\cite{ourdilaton,Chacko}. We will not consider the case of a Higgs-like dilaton pole at 125 GeV in this paper. This strong CFT sector is characterized by its $n$-point functions, which will be denoted by blobs in the diagrams below. The scale of conformal symmetry breaking will be parametrized by a single parameter $\mu$.

Within the SM one does not expect large deviations from mean-field theory in the electroweak symmetry breaking sector: any such deviation would be the result of small quantum loop corrections involving perturbative couplings. As a consequence, non-trivial momentum dependence in the SM is typically sub-dominant, leading to form factors that are constant up to small corrections. Once the effects of the strong sector leading to the QPT are added via its $n$-point functions, deviations from mean-field theory are possible. By now the Higgs boson has been observed in several different channels at the LHC, with all results in agreement with the SM predictions. When the Higgs is embedded into a strong conformal sector, one therefore needs to ensure that the resulting deviations from the SM are not too large. We first present a general parametrization of the relevant Higgs amplitudes in terms of form factors (Sections~\ref{sec:on-shell} and~\ref{sec:off-shell}), and then argue that most of these corrections remain small even when the Higgs is embedded into a strongly interacting sector (Sec.~\ref{sec:NDA}). We then explore specific realizations of the strong conformal sector and their contributions to these form factors in the later sections. We focus on those implementations where the corrections to already measured Higgs observables are expected to be small. The amplitudes can be divided into two sets: those where the measurement is made with all the legs on-shell (and the form factor reduces to a constant), and those measured with off-shell information, where additional momentum dependence is expected to appear.  

\subsection{On-shell behavior: Constant form factors\label{sec:on-shell}}

The majority of studies that have come out of the LHC are on properties of Higgs-pole observables, which implies that for most of the analyses  factorization can be employed to simplify theoretical predictions.  In this work, we restrict to $CP$-symmetric QPTs where the Higgs emerges as a $CP$-even state. Below we will be using diagrammatic illustrations for the form factors parametrized and estimated in this section. The dashed lines always represent Higgs-like states, either pole or continuum, while the shaded blob with several dashed lines represents an $n$-point function of the CFT. Multiple dashed lines stand for an arbitrary number of scalar insertions connecting the CFT with the external elementary legs. A cross will imply mixing between elementary and CFT states, while wiggly and straight lines represent gauge fields and fermions as usual.

For Higgs decays to fermions (\ie $\bar{b} b$ and $\bar{\tau} \tau$), the form factor for the coupling $hff $ is given by
\begin{center}
\includegraphics{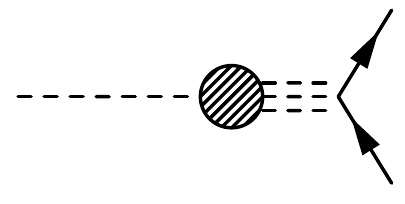}
\end{center}
\begin{equation}
 {\mathcal M}_{\bar{f} f h} = \bar{u}_1^a F_{hff} (p_1 \cdot p_2;\mu) v_{2a}~,
\end{equation}
where $\mu$ represents the parametric dependence on the scale of conformal symmetry breaking, and $p_{1,2}$ are the four-momenta of the two external fermions. In general a form factor involving $n$ fields depends on $n-1$ four-momenta,   which correspond to $n(n-1)/2$ Lorentz invariants. With the three external particles on-shell, the three Lorentz invariants are completely fixed, hence this form factor is simply an effective coupling constant, with $p_1 \cdot p_2 = m_h^2/2$.
 
For production of the Higgs via gluon fusion, there are modifications due to a non-perturbative sector that is coupled to the top quark:
\begin{center}
\includegraphics{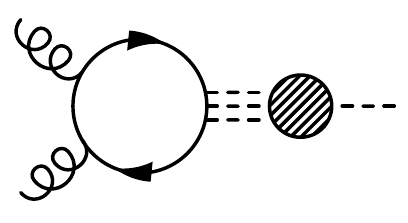}
\end{center}
\begin{equation}
{\mathcal M}_{ggh} = \left[ \left( \epsilon_1 \cdot p_2 \right) \left( \epsilon_2 \cdot p_1 \right) - \frac{m_h^2}{2} \left( \epsilon_1 \cdot \epsilon_2 \right) \right] F_{gg h} \left( p_1 \cdot p_2 = m_h^2/2 ; \mu \right)
\label{gghonshell}
\end{equation}
where $p_{1,2}$ are the (incoming) four-momenta of the two external gluons, and $\epsilon_{1,2}$ are their polarization vectors. 
The restriction to on-shell external states for the Higgs and the gluons implies that the form factor is simply an effective coupling constant.

There is an analogous formula corresponding to the form factor for the interactions of the Higgs with two photons. However, several contributions, such as the loop of vector bosons, change its value relative to the $hgg$ factor:
\begin{center}
\includegraphics{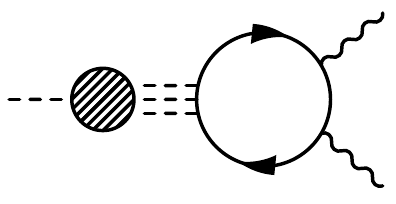} \hspace{0.5in} \includegraphics{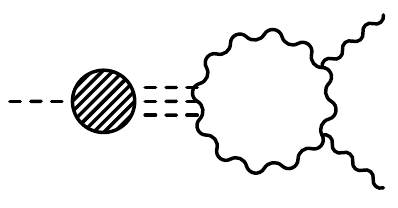} \hspace{0.5in} \includegraphics{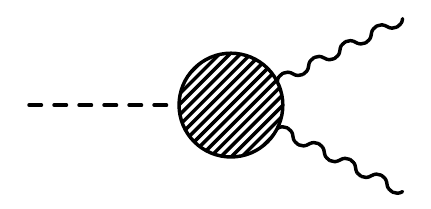} 
\end{center}
\begin{equation}
{\mathcal M}_{\gamma\gamma h} = \left[ \left( \bar{\epsilon}_1 \cdot p_2 \right) \left( \bar{\epsilon}_2 \cdot p_1 \right) - \frac{m_h^2}{2} \left(\bar{ \epsilon}_1 \cdot \bar{\epsilon}_2 \right) \right] F_{\gamma\gamma h} \left( p_1 \cdot p_2 = m_h^2/2 ; \mu \right)
\end{equation}
Of particular interest is the last class of diagrams, where electrically charged states in the strong sector contribute to the low-energy $h\gamma\gamma$ interaction.  This type of contribution is always present when there are states in the non-perturbative sector carrying charges under the SM $SU(2)_L \times U(1)_Y$ gauge group, like when the Higgs doublet operator mixes with or emerges from the strong dynamics.  Once again, restriction to on-shell states reduces this form factor to an effective coupling constant.

\subsection{Off-shell behavior: Momentum-dependent form factors\label{sec:off-shell}}

The previous form factors encode information about the new dynamics beyond the SM. Scaling dimensions of operators and $n$-point correlators in the strongly coupled sector enter into the corrections to the effective coupling constants, which can be measured at collider experiments.  However, in order to fully probe the nature of the quantum critical point we need to uncover the momentum dependence of the form factors, \ie when the Higgs is off-shell. Next we will outline the general structure of these off-shell observables.

The amplitude associated with the production of the Higgs through the fusion of massive vector bosons (VBF) has five independent contributions:
\begin{center}
\includegraphics[width=.3\textwidth]{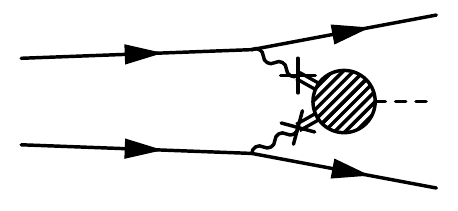}
\end{center}
\beq
{\mathcal M}_{VBF} \!\!\!&=&\!\!\!   J_1^\alpha G^{V}_{\alpha\mu}(p_1) \ J_2^\beta G^{V}_{\nu\beta}(p_2) F^{\mu\nu}_{VVh} \left( p_i;\mu \right) N_{V}
\label{VBFformfactors}\\
F^{\mu\nu}_{VVh} \left( p_i;\mu \right) \!\!\!&= &\!\!\! g^{\mu\nu}\, \Gamma_{1} +\left(g^{\mu\nu}p_1\cdot p_2- p_2^\mu p_1^\nu\right)\,\Gamma_{2} +\left(p_1^\mu p_1^\nu+p_2^\mu p_2^\nu \right)\, \Gamma_{3} +\left(p_1^\mu p_1^\nu -p_2^\mu p_2^\nu \right)\,\Gamma_{4} + \, p_1^\mu p_2^\nu\, \Gamma_{5} \, , \nonumber \\ 
&& \Gamma_i = \Gamma_i(p_1^2,p_2^2,p_1\cdot  p_2) \, ,
\eeq
where $p_{1,2}$ are the (incoming) four-momenta of the two vector bosons $V=\{W,\,Z\}$, and the on-shell Higgs condition fixes $p_1^2 + p_2^2 = m_h^2 - 2 p_1 \cdot p_2$. The fermionic currents $J_{1,2}$ contain the polarization states for the external fermions, with the appropriate Dirac structures that couple to the internal $W$ and $Z$ propagators, $G^{V}(p_i)$. $N_{V}$ is an overall normalization set to the SM value. Because of Bose symmetry, the $\Gamma_i$ form factors are symmetric under the exchange $p_1\leftrightarrow p_2$, except $\Gamma_4$ which is anti-symmetric. 
The gauge bosons' propagators include contributions from the CFT, which are determined to leading order once the Higgs two-point function is known.
Typically the $W$'s or $Z$'s are off-shell; the on-shell limit (a.k.a.~the effective $W$ limit) is relevant for high momenta, where the Higgs is far off shell.  
The $\Gamma_1$ form factor is the only one that appears at tree-level in the SM, where $\Gamma^{(\mathrm{SM})}_1=1$ and $\Gamma^{(\mathrm{SM})}_{i\neq1}=0$. The form factor $\Gamma_2$ is singled out as the term transverse to the momenta $p_1$ and $p_2$, thus it is generated by operators involving transverse polarizations.

For Higgsstrahlung, where an off-shell electroweak vector boson produced via Drell-Yan radiates an on-shell Higgs and a massive vector, the amplitude involves the same form factors (up to $p_2 \rightarrow -p_2$) as VBF, since the two processes are related by crossing symmetry:
\begin{center}
\includegraphics{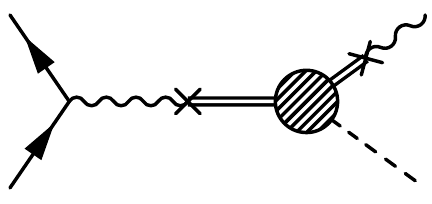}
\end{center}
\beq
{\mathcal M}_{\bar{q} q \rightarrow V h}  =   J_1^\alpha G^V_{\alpha\mu}(p_1)\, \bar{\epsilon}_{2\,\nu}   F^{\mu\nu}_{VVh}\left(p_1,-p_2;\mu \right) N_{V} 
 \eeq
where $J_1$ is the current of initial state fermions, $\epsilon_2$ is the vector boson polarization vector, and 
\beq
\bar{\epsilon}_{2\,\nu} F^{\mu\nu}_{VVh} \left(p_1,-p_2;\mu \right) \!\!\!&=&\!\!\!  \bar{\epsilon}_{2}^{\, \mu} \Gamma_{1} - \left[ \bar{\epsilon}_{2}^{\, \mu}  \, p_1\cdot p_2 - p_2^\mu \, \bar \epsilon_2\cdot p_1 \right] \,\Gamma_{2} + p_1^\mu \, \bar \epsilon_2\cdot p_1 \, \left( \Gamma_{3}+\Gamma_{4} \right) \, , \nonumber \\
&& \Gamma_{i} = \Gamma_{i}(p_1^2, p_2^2, -p_1\cdot  p_2) \, ,
\eeq
where we used $\epsilon_2\cdot p_2=0$. Therefore, Higgsstrahlung contains three form factors, in agreement with \cite{Isidori}. Note that $p_2^2 = m_V^2$ and $-2 p_1\cdot  p_2 = m_h^2 - m_V^2 - p_1^2$.

Double Higgs production involves two  form factors:
\begin{center}
\includegraphics[width=2in]{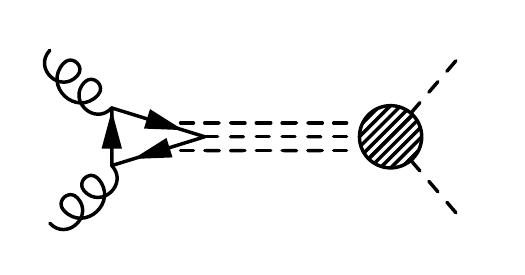}
\end{center}
\beq
{\mathcal M}_{gghh} \!\!\!&=&\!\!\! \big[ \left( \epsilon_1 \cdot p_2 \right) \left( \epsilon_2 \cdot p_1 \right) - (p_1\cdot p_2)
\left( \epsilon_1 \cdot \epsilon_2 \right) \big] \, \Xi_1 \left( p_1 \cdot p_2, p_1 \cdot p_3 ; \mu \right) \nonumber \\
&&\!\!\! + \epsilon_2 \cdot \left[ (p_1\cdot p_2) p_3 - (p_2\cdot p_3)p_1\right] \, \epsilon_1 \cdot\left[(p_1\cdot p_2) p_3 -  (p_1\cdot p_3) p_2 \right] \, \Xi_2 \left( p_1 \cdot p_2, p_1 \cdot p_3 ; \mu \right)
\eeq
where $p_{1,2}$ and $p_{3,4}$ are respectively the four-momenta of the two gluons (incoming) and of the two Higgses (outgoing). Note that $p_1 \cdot p_2 = s/2$ and $p_1 \cdot p_3 = (m_h^2-t)/2$. Because of Bose symmetry between the gluons,  $\Xi_i \left( p_1 \cdot p_2, p_1 \cdot p_3 ; \mu \right)=\Xi_i \left( p_1 \cdot p_2, p_2 \cdot p_3 ; \mu \right)$. 
The two different structures are generated already in the SM, although $\Xi_2$ is suppressed in the large top mass limit \cite{Plehn:1996wb}. 

The last and perhaps most promising case is $gg\rightarrow VV$, in particular when $V = Z \rightarrow \ell^+ \ell^-$:
\begin{center}
\includegraphics[width=2in]{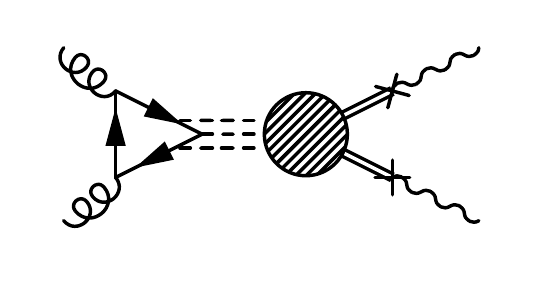}\nonumber
\end{center}
\beq
{\mathcal M}_{ggVV} = \epsilon_{1\, \mu} \epsilon_{2\, \nu} \left[ F^{\mu\nu\rho\sigma}_{ggVV} \left( p_i;\mu \right) + \widehat{F}^{\mu\nu\rho\sigma}_{ggVV}\left( p_i;\mu \right) \right] \bar{\epsilon}_{3 \, \rho} \bar{\epsilon}_{4 \, \sigma}
\label{ggVVformfactors}
\eeq
\beq
F^{\mu\nu\rho\sigma}_{ggVV} \left( p_i;\mu \right) \!\!\!&= &\!\!\!
\big[ g^{\mu \nu} (p_1 \cdot p_2) - p_1^\nu p_2^\mu \big] 
\big( g^{\rho \sigma} \Theta_1 + p_1^\rho p_1^\sigma \Theta_2 + p_2^\rho p_2^\sigma \Theta_3 \big) \\
&&\!\!\! + \, \big[ g^{\mu \rho} g^{\nu \sigma} (p_1 \cdot p_2) + g^{\mu \nu} p_1^\rho p_2^\sigma - g^{\mu \rho} p_1^\nu p_2^\sigma - g^{\nu \sigma} p_2^\mu p_1^\rho \big] \, \Theta_4 \nonumber \\
&&\!\!\! + \, g^{\rho \sigma} \big[ g^{\mu \nu} (p_1 \cdot p_3) (p_2 \cdot p_3) - p_3^\mu p_3^\nu (p_1 \cdot p_2) + p_3^\mu p_1^\nu (p_2 \cdot p_3) + p_2^\mu p_3^\nu (p_1 \cdot p_3) \big] \, \Theta_5 \nonumber \\
&&\!\!\! + \, p_3^\sigma \big[ g^{\mu \nu} p_2^\rho (p_1 \cdot p_3) + g^{\nu \rho} p_3^\mu (p_1 \cdot p_2) - g^{\nu \rho} p_2^\mu (p_1 \cdot p_3) - p_3^\mu p_1^\nu p_2^\rho \big] \, \Theta_6 \nonumber \\
&&\!\!\! + \, \big[ g^{\mu \sigma} p_1^\nu p_1^\rho (p_2 \cdot p_3) - g^{\mu \rho} p_1^\nu p_1^\sigma (p_2 \cdot p_3) + g^{\mu \rho} p_1^\sigma p_3^\nu (p_1 \cdot p_2) - g^{\mu \sigma} p_1^\rho p_3^\nu (p_1 \cdot p_2) \big] \, \Theta_7 \nonumber \\
&&\!\!\! + \, \big[ g^{\nu \sigma} p_2^\mu p_2^\rho (p_1 \cdot p_3) - g^{\nu \rho} p_2^\mu p_2^\sigma (p_1 \cdot p_3) + g^{\mu \rho} p_1^\sigma p_3^\nu (p_1 \cdot p_2) - g^{\mu \sigma} p_1^\rho p_3^\nu (p_1 \cdot p_2) \big] \, \Theta_8 \, , \nonumber \\
\widehat{F}^{\mu\nu\rho\sigma}_{ggVV} \left( p_i;\mu \right) \!\!\!&= &\!\!\! p_{1\,\alpha} p_{2\,\beta} p_{i\,\gamma} p_{j\,\delta} \left( \varepsilon^{\mu\nu\alpha\beta} \varepsilon^{\rho\sigma\gamma\delta} \widehat{\Theta}^{ij}_1 +
 \varepsilon^{\mu\rho\alpha\gamma} \varepsilon^{\nu\sigma\beta\delta} \widehat{\Theta}^{ij}_2 
+  \varepsilon^{\mu\sigma\alpha\gamma} \varepsilon^{\nu\rho\beta\delta} \widehat{\Theta}^{ij}_3 \right. \\
&&\left. + \, \delta^i_1 \delta^j_3 \varepsilon^{\mu\alpha\rho\sigma} \varepsilon^{\nu\beta\gamma\delta} \widehat{\Theta}_4 + \delta^i_2 \delta^j_3 \varepsilon^{\nu\beta\rho\sigma} \varepsilon^{\mu\alpha\gamma\delta} \widehat{\Theta}_5  \right) \, , \nonumber\\
&&\!\!\! \Theta_{k} = \Theta_{k}(p_1 \cdot p_2, p_1 \cdot p_3)  \, , \quad \widehat{\Theta}^{ij}_{k} = \widehat{\Theta}^{ij}_{k}(p_1 \cdot p_2, p_1 \cdot p_3)  \nonumber
\eeq
where $p_{1,2}$ are the four-momenta of the two gluons, with polarizations $\epsilon_{1,2}$, and $p_{3,4}$ are the four-momenta of the two massive electroweak vectors, with $\epsilon_{3,4}$ their polarization. Note that $p_1 \cdot p_2 = s/2$ and $p_1 \cdot p_3 = (m_V^2-t)/2$. The latin indices run over a set of three independent particle momenta (e.g.~1 to 3). We have used a slightly redundant notation: some of the form factors vanish identically because of the anti-symmetry of the $\epsilon$ tensor, $\widehat{\Theta}^{1j}_{2,3}=\widehat{\Theta}^{i2}_{2,3}=0$; moreover, because of Bose symmetry between the gluons,  $\widehat{\Theta}_{3,5}$ can be expressed in terms of $\widehat{\Theta}_{2,4}$ respectively. More explicitly, $\widehat{\Theta}^{ij}_1(p_1\cdot p_2, p_1\cdot p_3)=\widehat{\Theta}^{ij}_1(p_1\cdot p_2, p_2\cdot p_3)$, $\widehat{\Theta}^{ij}_{2}(p_1\cdot p_2, p_1\cdot p_3)=\widehat{\Theta}^{ji}_{3}(p_1\cdot p_2, p_2\cdot p_3)$ and $\widehat{\Theta}_4(p_1\cdot p_2, p_1\cdot p_3)=\widehat{\Theta}_5(p_1\cdot p_2, p_2\cdot p_3)$. Analogous  Bose symmetry constraints apply to other $\Theta$'s as well. 
For recent analyses of this process at the LHC, in the limit $p^2 \ll \mu^2$ where the resulting effective theory is the SM perturbed by higher-dimensional operators, see e.g.~\cite{ggtozzeft}.

Eventually, with a large integrated luminosity, we will be able to probe $VV$ scattering, $V=\{W,\,Z\}$: 
\beq
\includegraphics[width=2in]{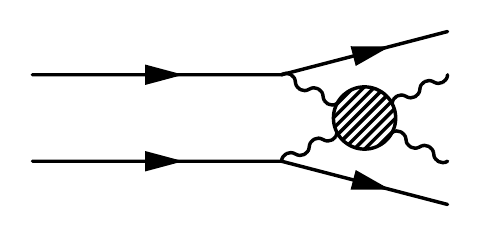}\nonumber
\eeq
In general there are many form factors in this channel, even restricting to on-shell $V$'s. Likewise for the Higgs associated production with a top quark pair:
\begin{center}
\includegraphics{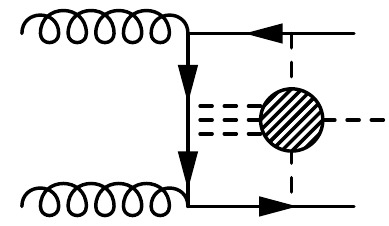}
\hspace{1cm}
\includegraphics{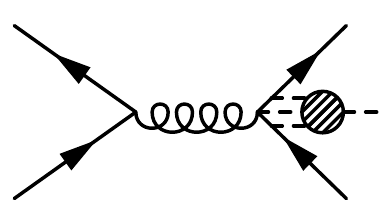}
\end{center}
either $gg$ or $q\bar{q}$ initiated. We leave the analysis of these form factors for future work. 

\subsection{Estimation of form factors: A case for a simple parametrization of the leading Higgs processes\label{sec:NDA}}

Next we will argue that most form factors remain small and can be estimated using the insertions of the smallest $n$-point functions, even though the Higgs is (partly) embedded into a strongly coupled sector.
In the framework of a QCH we assume that the origin of the corrections to Higgs observables is from a strongly coupled sector that plays some role in electroweak symmetry breaking, and perhaps produces the light Higgs boson resonance (possibly along with a scalar continuum, as we will discuss later). In this paper we assume all other SM fields are external to this non-perturbative sector, with the Higgs acting as a portal to the strong dynamics. Naive dimensional analysis (NDA) provides a guide for our expectations of the size of the form factors discussed above. We begin our estimates from an effective field theory perspective, where the only light degree of freedom surviving from the strongly coupled sector (below the scale $\mu$) is the Higgs particle.  In this limit, the form factors are expected to have asymptoted to their zero-momentum values~\footnote{Departures from this limit scale with $p^2/\mu^2$, where $p$ is the typical momenta of the process and $\mu$ the mass scale of the heavy degrees of freedom that have been integrated out.}. In this case, we can estimate the size of the $n$-point Higgs correlators by considering the effect of loops on its renormalization. In NDA the loop corrections should be roughly of the same size as the original $n$-point function. The example of $n=6$ is explicitly depicted here:
\begin{center}
\includegraphics[width=2.5in]{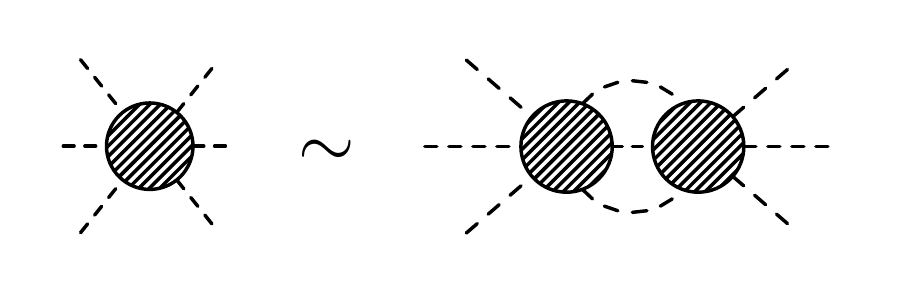}
\end{center}
Following the rules of NDA, assuming that the $n$-point function is described by the coupling
\begin{equation}
{\mathcal  L} = \frac{\alpha_n}{\mu^{n-4}} \phi^n \, ,
\label{npoint}
\end{equation}
a typical loop contribution with two insertions of this operator that contributes to the same $n$-point amplitude would be one in which each vertex has $n/2$ external lines, and $n/2$ propagators exchanged in $n/2-1$ loops that are cut off at the scale $\mu$. In this case, the quantum correction to $\alpha_n$ is expected to be roughly
\begin{equation}
\alpha_n \rightarrow \alpha_n  \left( 1 + \frac{\alpha_n}{(16 \pi^2 )^{n/2-1}} \right) \, . \nonumber
\end{equation}
For this correction to be comparable to the initial coupling we must have $\alpha_n \sim (16 \pi^2)^{n/2-1}$~\footnote{Large-$N$ theories may provide extra $1/N$ suppression to the connected higher-$n$ point functions.}.

With this NDA estimate of the $n$-point amplitude in hand, we can see that the $n$-point contribution to, e.g., the gluon fusion process is suppressed by insertions of the perturbative coupling of the top-quark to the strongly coupled sector, along with a loop factor that is only partially cancelled by the large coefficient $\alpha$. 
\begin{center}
\includegraphics[width=2.5in]{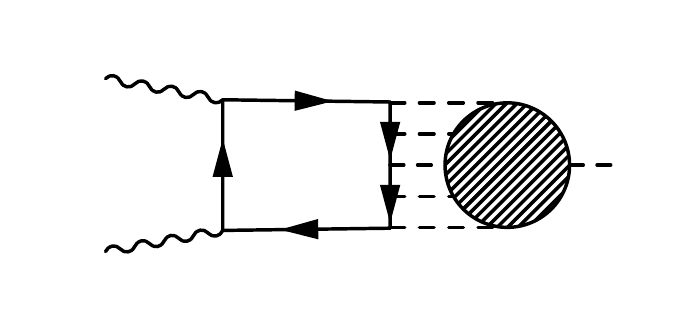}
\end{center}
If the shaded region corresponds to the $n$-point function, there are $n-1$ insertions of the top Yukawa, and $n-2$ loops.  There are $n-1$ scalar propagators and $n-2$ fermionic propagators running in these loops.   Computing the loops with a hard cutoff at the scale $\mu$ in \eq{npoint}, yields an estimate for the contribution of the $n$-point correlator to the $h \bar{t} t$ coupling:
\begin{equation}
g^{tth}_n \sim 4 \pi \left( \frac{\lambda_t}{4\pi} \right)^{n-1}.
\end{equation}
This is one of the crucial results in this paper: one that allows us to use very simple parametrizations to estimate the leading corrections to Higgs processes in QCH models. 
This NDA does not rely on whether or not the strong sector is conformal.
What this means is that the $n$-point amplitude contributions are increasingly suppressed by perturbative loop factors for increasing $n$, and therefore the leading contribution to the form factor in this case will be due to the Higgs two-point function (in the electroweak broken phase). In this case, the dominant contribution to the form factor is ``tree-level", and involves only a single insertion of the top Yukawa coupling along with the full non-perturbative two-point function of the Higgs. For example, non-standard momentum-dependent effects in double Higgs production through gluon fusion would be dominated by the following diagram
\begin{center}
\includegraphics[width=2.5in]{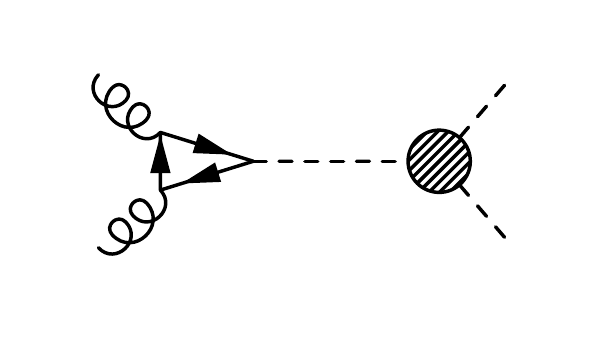}
\end{center}
which involves the $ggh$ form factor in \eq{gghonshell}, now with the Higgs off-shell ($p_1 \cdot p_2 = s/2$), and the trilinear Higgs form factor $F_{hhh}$:
\beq
\label{gghhNDA}
{\mathcal M}_{gghh} = \left[ \left( \epsilon_1 \cdot p_2 \right) \left( \epsilon_2 \cdot p_1 \right) - (p_1 \cdot p_2) \left( \epsilon_1 \cdot \epsilon_2 \right) \right] F_{gg h} \left( p_1 \cdot p_2 ; \mu \right) \, G(p_1 + p_2) \, F_{hhh}(p_1 \cdot p_2;\mu) \, ,
\eeq
where $G(p_1 + p_2)$ is the Higgs propagator.
 We emphasize again that this result is based on the crucial (but reasonable) assumption of the applicability of our NDA to estimate contributions from the strong sector.

The same power counting implies that the form factor $F_{gg h}$ with the Higgs on-shell reduces to the SM value up to corrections of $O(\lambda_t/4\pi)$ associated with higher-point correlation functions, and up to corrections of $O(m_h^2/\mu^2)$ due to the non-trivial momentum dependence of such  higher-point correlators.

Analogously, the strong sector's contribution to the $gg\rightarrow VV$ amplitude is dominated by the following diagram
\begin{center}
\includegraphics[width=2in]{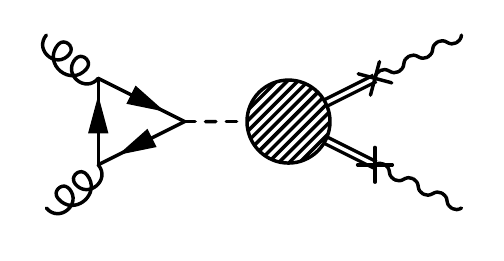}
\end{center}
with the associated form factor:
\beq
{\mathcal M}_{ggVV} = \left[ \left( \epsilon_1 \cdot p_2 \right) \left( \epsilon_2 \cdot p_1 \right) - (p_1 \cdot p_2) \left( \epsilon_1 \cdot \epsilon_2 \right) \right] F_{gg h} \left( p_1 \cdot p_2 ; \mu \right) \, G(p_1 + p_2) \, 
F^{\mu\nu}_{VVh} \left(-p_3, -p_4;\mu \right) \bar{\epsilon}_{3\,\mu} \bar{\epsilon}_{4\,\nu}\,.
\eeq
Finally, the new physics effects in $VV$ scattering factorize into the product of the VBF form factors. The Higgs contribution to $VV$ scattering is singled out, even off-shell, and the $s$-channel Higgs amplitude is given by
\beq
{\mathcal M}_{WWWW} = J_1^\mu G_{\mu\nu}(p_1) J_2^\rho G_{\rho\sigma}(p_2)  F^{\nu\sigma}_{VVh}\left(p_1, p_2;\mu \right) \, G(p_1+p_2) \, F^{\alpha \tau}_{VVh}\left(-p_3,-p_4;\mu \right) \bar{\epsilon}_{3 \, \alpha} \bar{\epsilon}_{4 \, \tau} \, .
\label{WWformfactors}
\eeq
This probes both the Higgs propagator and the two gauge boson form factors off-shell.

Let us comment on the UV scaling behavior of the form factors for the case of a strong sector that is conformal at high energies. For large momenta, the Higgs propagator scales as $p^{2\Delta-4}$, where $\Delta$ is the dimension of the Higgs-operator. The amputated form factors $F_{h\ldots h}$ scale asymptotically as $p^{4-n\Delta} \cdot v^{m\Delta}$, where $n$ is the number of legs with large momenta while $m$ is the number of VEV insertions at zero momentum. For example, $F_{hhh}$ in \eq{gghhNDA} requires one VEV and three Higgs legs, so that $F_{hhh} \sim p^{4-3\Delta}v^\Delta$. Analogously, the form factor $F^{\mu\nu}_{VVh} \epsilon_{\mu} \epsilon_{\nu}$ scales at large momenta as $F_{hhh}$ because of the equivalence theorem that relates the Goldstone bosons inside the Higgs field to the longitudinal polarizations of $V$. For the transverse polarizations, the estimate of the asymptotic behavior involves instead the insertion of the weakly gauged conserved currents of the CFT, schematically $\langle H^\dagger\ldots H J_\mu J_\nu\rangle$, so that the scaling of the amputated form factor is $p^{-n\Delta+2}\cdot v^{m\Delta}$. 
We stress that our power counting implicitly assumes that the IR corrections from insertions of several Higgs VEV's are suppressed, such that the $n$-point function with the least number of critical Higgses (but with unsuppressed number of derivatives) captures already the leading contribution to the form factors. Such a scaling is realized e.g.~in the power counting of \cite{Liu:2016idz}, where operators with fewer derivates are suppressed due to a shift symmetry $h\rightarrow h+c$.
 \\

We emphasize that these results depend crucially on how the SM fields are embedded into the (strongly interacting) conformal sector. If the top quark were part of the strong dynamics, as it is often the case in extensions of the standard model, then the higher-point correlation functions are expected to give contributions that are unsuppressed relative to the leading term in the $\lambda_t$ expansion.  In this case, all $n$-point correlators are important for estimating the form factors.
In fact, this is the case for the longitudinal $W^\pm$ bosons, which are part of the strong sector (along with the Higgs and the longitudinal $Z$) and may hence strongly affect the rate of $h\rightarrow \gamma\gamma$.  Of course, the argument above is model dependent. For example, in composite Higgs models where the Higgs emerges as a Goldstone boson, corrections to $h\rightarrow \gamma\gamma$ (and $gg\rightarrow h$) are protected by the associated Goldstone symmetry. Such a symmetry is preserved by the strong sector and therefore the generation of terms such as $B_{\mu\nu}^2 |H|^2$ (and $G_{\mu\nu}^2|H|^2$) is suppressed by insertions of the explicit breaking parameters \cite{Giudice:2007fh}. 
In the next section we present a \emph{different} class of strongly interacting theories that can reproduce the SM predictions at low energies, e.g.~in $h\rightarrow \gamma\gamma$, even though no global symmetry is at work. Rather than invoking a symmetry, we consider generic theories that are perturbations around generalized free fields \cite{Greenberg:1961mr}.  For example, below we consider a strong sector where only the $n=2$ correlators are non-vanishing, and where higher-point correlators are present only due to the perturbative SM couplings.  Another example we consider is that in which higher $n$-point correlators are suppressed due to a large $N$ expansion being possible in the strong sector.  In either case, these benchmark theories admit a perturbative expansion, where the smallness of the SM couplings or of $1/N$ suppresses deviations from the new physics sector. Effectively, the scale suppressing operators associated with higher $n$-point functions is larger than the scale that controls the two-point function. Hereafter, this latter scale alone will be denoted as $\mu$.

\section{Modeling the Quantum Critical Higgs}

The range of possible phenomena associated with generic models of electroweak quantum criticality is large, with only a few constraints imposed by consistency of the theory (e.g.~unitarity bounds) on the form factors discussed above.  To make concrete predictions we need to make some additional assumptions about the QPT.  In this section we will present two such models: one based on generalized free fields and a second one based on a 5D realization of those using the AdS/CFT correspondence. These models serve as illustrations of the general arguments presented in Sec.~\ref{sec:NDA}, and serve as toy models for performing concrete calculations in the upcoming section for LHC phenomenology. While not forbidden by experimental results, strong self-interactions in the low-energy effective theory for the Higgs would hinder our ability to make quantitative statements. Of course there is no requirement that strongly coupled dynamics produces a strongly coupled effective theory, and in fact there are many counter examples where a strongly coupled theory produces a weakly coupled effective theory for the low-energy degrees of freedom: Seiberg duality, the $\rho$ meson at large $N$, and the AdS/CFT correspondence. In the latter example, assuming that the low-energy composites of the broken CFT have weak interactions is tantamount to assuming that there is a weakly coupled AdS dual.  In the strict large $N$ limit there are no bulk interactions in the AdS dual, but for large yet finite $N$ we expect that there are perturbative bulk interactions that are inherited by the 4D low-energy effective theory.  

The infinite $N$ limit in AdS/CFT yields a subclass of models that generalize to a broader category of strongly coupled theories:  it is possible that the strongly coupled sector is completely specified by the two-point functions of that theory, with or without a large $N$ expansion.  Such theories are referred to as models of generalized free fields \cite{Greenberg:1961mr}.  Weakly coupling a fundamental light Higgs to such a theory would produce the type of dynamics we are interested in.  Of course, such a construction would not on its own resolve the hierarchy problem, however our motivation is to explore the possible variations of Higgs phenomenology rather than solving the hierarchy problem.

\subsection{Generalized free field theory}
\label{sec:GFFT}

In theories where $n$-point functions with $n > 2$ vanish, one obtains what is called a ``generalized free field theory" \cite{Greenberg:1961mr}.  For a more recent discussion of generalized free fields see e.g.~\cite{gffmodern} and references therein. Since the theory is quadratic in this case, a 1PI effective Lagrangian density can be constructed whose path integral generates the two-point functions of the theory.  As an  example, we consider an unbroken CFT with a scalar operator $\Qh$ with scaling dimension $\Delta$. The two-point function is then fixed by conformal invariance:\footnote{This is also the unparticle propagator of Georgi \cite{Georgi}.}
\beq
G(p^2) = -\frac{i}{\left(-p^2+ i \epsilon \right)^{2-\Delta}} \, .
\eeq
The Lagrangian that reproduces this two-point function is
\begin{equation}
{\mathcal L}_\text{GFF} = -\Qh^\dagger \left(\partial^2\right)^{2-\Delta} \Qh \, .
\end{equation}

Phenomenological constraints suggest that if there is a strongly coupled sector mixing with the Higgs, then there must either be a gap, or the mixing must be highly suppressed.  A simple IR deformation of the above Lagrangian provides a two-point function that features a gap, yet reduces to conformal behavior at high momentum:
\begin{equation}
{\mathcal L} = -\Qh^\dagger \left(\partial^2+ \mu^2 \right)^{2-\Delta} \Qh \, .
\label{quadraticaction}
\end{equation}
Here $\mu$ reduces to a mass term (a pole in the two-point function) as $\Delta$ goes to 1, but for other values of $\Delta$ represents the beginning of a cut. The $\mu$ term gives a contribution to the potential energy (the $p\to 0$ limit) and removes the massless degrees of freedom. In terms of a fundamental CFT description this would correspond to the continuum shifted to start at $p^2=\mu^2$ rather than at $p^2=0$.  

There are other possibilities \cite{other,PerezVictoria:2008pd} for the structure of the above quadratic Lagrangian that correspond to differently shaped spectral density functions.  The different shapes correspond to different ways in which the behavior of the theory makes the transition from the IR, where conformal symmetry is broken, to the UV, where it is restored. For the sake of simplicity we will use examples based on this simple model, but want to emphasize again that this is not a necessary or unique choice. The appearance of a continuum in conformal theories is generic; such a CFT continuum does not admit, generically, an interpretation in terms of weakly interacting multiparticle states.  Whether or not this continuum survives the spontaneous breaking of the conformal symmetry (or becomes a discretuum or a continuum with a mass gap) depends strongly on the mechanism of conformality breaking and the corresponding CFT dynamics. We will see in the section below that for CFT's with AdS duals a continuum theory generically corresponds to soft-wall type constructions, even though soft walls may also support a mass gap.  Hard walls correspond to a discretuum, as in the original RS models, although finite coupling quantum effects in the 5D theory restore the continuum at high scales (where the mode separation becomes small in comparison with the widths).

In general, the two-point function can be formulated in terms of a spectral density function.  Of course, with the discovery of the Higgs particle, we insist that the spectral density includes at least a pole at $125$~GeV, with features that closely resemble those of the SM Higgs.  A general two-point function with a pole at $m_h^2$, and a cut beginning at the scale $\mu$ is
\begin{equation}
G_\Qh (p^2) = \frac{i}{p^2-m_h^2} + \int_{\mu^2}^\infty dM^2 \frac{ \rho (M^2)}{p^2-M^2} \, .
\end{equation}
A simple Lagrangian that yields a two-point function of the above form is
\begin{equation}
{\cal L}_\text{quadratic} = 
-\frac{1}{2\, Z_\Qh} \Qh \left[ \partial^2+\mu^2\right]^{2-\Delta} \Qh + \frac{1}{2\,Z_\Qh} (\mu^2-m_h^2)^{2-\Delta} \Qh^2~,
\label{qptlag}
\end{equation}
where the physical pole mass, $m_h \approx 125$ GeV, and the quadratic terms in \eq{qptlag} are obtained by expanding the Higgs potential around its minimum, such that $\langle \Qh\rangle=0$, $\Qh$ being the real part of the fluctuation around the VEV. 
Note that while the scaling dimension of $\Qh$ is $\Delta$, its engineering dimension is 1. Since $\Qh$ is assumed to be weakly coupled, the scaling dimension of $\Qh^2$ is $2 \Delta$ to first order, so the bounds of \cite{Vichi:2011ux} are not relevant. Perturbative corrections will also give additional contributions to the spectral density, shifting both the location of the branch cut, and the overall shape.  These effects, though, are both sub-dominant, and we neglect them in our estimation of form factors.
It is convenient to work with a canonically normalized Higgs field.  To achieve this we can set the residue of the pole to 1 by choosing the normalization 
to be
\begin{equation}
Z_\Qh = \frac{\left(2-\Delta\right)}{\left(\mu^2-m_h^2\right)^{\Delta -1}} \, .
\end{equation}
The propagator for the physical Higgs scalar can then be written simply as
\beq
\label{propagator}
G_\Qh(p)=-\frac{i \,Z_\Qh}{(\mu^2-p^2+i\epsilon)^{2-\Delta} -(\mu^2-m_h^2)^{2-\Delta}} \,.
\eeq
This type of propagator has been studied in a variety of papers \cite{Unhiggs,Irvine,coloredunparticles}, including its AdS$_5$ description \cite{AdSCFTUnP,Falkowski}.

Besides modified propagators, the Higgs Lagrangian \eq{qptlag} is associated with non-standard couplings between the Higgs field components and the transverse polarizations of the electroweak gauge bosons, once we promote the derivative in \eq{quadraticaction} to a gauge covariant derivative. In fact, it also gives rise to new vertices with arbitrary powers of the gauge fields. Here we will use the results of \cite{Unhiggs} which used the Mandelstam technique \cite{Mandelstam} of path-ordered exponentials (aka Wilson lines) to ensure gauge invariance.  The form factors that describe the interactions of the rescaled Higgs field $\Qh$ with gauge fields are given in Appendix A, along with further details on the Lagrangian \eq{qptlag}.

While a detailed derivation of the bounds on the parameters $\mu ,\Delta$ is beyond the scope of this paper, one can get a good idea of how weakly constrained these parameters are by going into the limit of large $\mu$ compared to the momentum scales relevant for a particular process, and expand the Lagrangian \eq{qptlag} in powers of $p^2/\mu^2$. The leading operator (after rescaling the Higgs doublet to have a canonical kinetic term) will be a dimension six Higgs operator $-\frac{1}{2} \frac{(1-\Delta)}{\mu^2} (D^2 H^\dagger) (D^2 H)$, which as expected vanishes both for $\Delta \to 1$ and $\mu \to \infty$. Using the equations of motion (or field redefinitions) we can gain more insight into the effects of this operator: it will induce 4-Fermi operators, which are however strongly suppressed by the SM Yukawa couplings, a modification to the Higgs potential (which is very weakly constrained by current data), while the leading effect will be a modification to the Yukawa couplings given by 
\begin{equation}
\frac{1}{2} \frac{(\Delta-1)}{\mu^2} \frac{Y}{\sqrt{2-\Delta}}   \bar{\psi}_L \frac{\partial V_H}{\partial H^\dagger} \psi_R + h.c. \, , 
\end{equation}
where $Y$ are the Yukawa couplings and $V_H$ is the Higgs potential. This will then give rise to a correction to the top Yukawa coupling, which will in turn modify the Higgs production rate via gluon fusion. 
The resulting correction (expressed in terms of the physical Higgs mass) together with the experimental limit at 1$\sigma$ CL, obtained from \cite{ATLASCMS} assuming no direct new physics contributions to the Higgs coupling to gluons, is given by 
\begin{equation}
\frac{-\delta y_{\psi \psi h}}{(y_{\psi \psi h})_{SM}} \simeq \frac{\Delta-1}{2} \frac{m_h^2}{\mu^2} \approx 3 \% \left(\frac{\Delta-1}{0.6}\right) \left( \frac{0.4 \TeV}{\mu} \right)^2 \lesssim 7 \% \, .
\label{tophiggs}
\end{equation}
leading to an experimental bound on the parameters $\mu/\sqrt{\Delta-1} \gtrsim 335 \GeV$.

Going beyond the quadratic terms we can also include small Higgs self-interactions with their associated form factors, but these are model dependent in that they require information beyond simply the two-point function. The simplest models that provide this kind of detail come from the AdS/CFT correspondence, which we discuss in what follows.

\subsection{Generalized free fields and AdS/CFT}

If we insist that the dynamics is conformal in the UV, as in the examples above, then the high-momentum behavior of all such two-point functions is purely a function of momentum and the scaling dimensions of the fields.  In the IR, where the conformal dynamics is presumed to be broken to produce a gap, the specifics of the breaking determine the transition from SM-like mean-field behavior to exhibiting sensitivities to the scaling dimensions associated with the strongly coupled conformal theory.
 
 The AdS/CFT correspondence~\cite{AdSCFT} offers a framework where the strongly coupled CFT with generalized free field behavior, its perturbative coupling to the fundamental fields of the SM, and its breaking can all be understood in terms of a weakly coupled 5D theory. We consider a class of 5D models in which a scalar field carrying the same quantum numbers as the Higgs propagates in the bulk of the extra dimension.  5D gauge fields are required for consistency, such that local gauge transformations of the bulk Higgs are compensated for appropriately.  A soft-wall is included to truncate the extra dimension, producing a gap in the spectrum near the TeV scale.  The rest of the SM fields are taken to be localized on the UV brane.  In general, these can propagate in the bulk as well, but we take the simplifying assumption that only the minimal bulk field content be added to generate non-trivial behavior for the Higgs QPT.
 
 The 5D theory that we consider has the following action:
\begin{equation}
S = \int d^4xdz \sqrt{g} \left[ \left| D_M H \right|^2 - \frac{1}{4 g_{4}^2} W_{MN}^{a~2} - \phi(z) \left| H \right|^2 + {\mathcal L }_\text{int}(H)
\right] + \int d^4 x\, {\mathcal L }_\text{perturbative} \, .
\label{5Daction}
\end{equation}
An $SO(4)$ global symmetry is gauged in the bulk, introduced in order to preserve the custodial $SU(2)_L \times SU(2)_R$ symmetry \cite{custsym} of the SM.  Smaller groups are possible, but are difficult to reconcile with electroweak precision constraints without a large separation of scales. The electroweak singlet $\phi$ is a background field whose expectation value determines the bulk Higgs mass, and whose profile determines the properties of the soft-wall and associated gap.
The absence of ${\mathcal L }_\text{int}$, with terms higher than quadratic, in the  5D description would correspond to the generalized free field limit.  Their inclusion allows for form factors for non-trivial, $n>2$ point, correlation functions.

Via the AdS/CFT correspondence, this action, with a constant background field $\phi =m^2$ and neglecting ${\mathcal L }_\text{int}$, encodes the physics of a large $N$ $4$D strongly coupled CFT containing a scalar operator with a scaling dimension given by 
 \beq
 \Delta = 2 \pm \sqrt{4 +m^2 R^2} \, .
 \label{dimension}
 \eeq  The 5D gauge fields correspond to the global symmetries of the approximate CFT.  At a minimum, it must contain the global symmetries that are gauged in the SM, and phenomenological viability typically forces invariance under custodial $SU(2)_L \times SU(2)_R$. Since we are interested in fields with dimensions $\Delta < 2$, we need to choose boundary conditions~\cite{Klebanov:1999tb,AdSCFTUnP} that project out the solution with larger root in Eq.~(\ref{dimension}), which results in the boundary value ($H_0$) of the bulk field ($H$) playing the role of the 4D effective field rather than the source of the CFT operator as it does when $\Delta >2$.
 
For the metric, $g$, we presume the space is asymptotically AdS, with metric 
\beq
ds^2_\text{UV} \approx \left(\frac{R}{z}\right)^2 \left(\eta_{\mu\nu}dx^\mu dx^\nu-dz^2\right) \, ,
\eeq
 in a region of the space $z\sim R$.  Deviations from AdS grow with increasing $z$, forcing a finite size for the extra dimension and a resulting mass-gap for the 5D modes.  The precise details of these deformations of AdS determine the spectrum -- whether there is a discretum or a continuum, and the detailed shape of the spectral density. A simple classification of the characteristics of such spacetimes has been given in~\cite{Falkowski} for the case when the metric is modified by an additional overall soft-wall factor, $ds^2 = a(z)^2 \left(\eta_{\mu\nu}dx^\mu dx^\nu-dz^2\right)$, and a bulk Higgs potential $V(H)$ is included.
The Higgs spectrum is determined by the Schr\"odinger-type equation~\cite{Falkowski}
\begin{equation}
[-\partial_z^2 +\hat{V}(z) ] \Psi = p^2 \Psi
\end{equation}
where $\hat{V}(z) = \frac{3}{2} \frac{a''}{a} + \frac{3}{4} \frac{a'^2}{a^2} + \hat{M}^2 (z) $ with $\hat{M}^2 = a^2 R V''(H)$. It was found that the asymptotic behavior of the potential determines the qualitative features of the Higgs spectrum. If $\hat{V} (z) \to \infty$ for $z\to \infty$ there is a discretuum, which is the case for all hard walls as well as soft walls where the warp factor decays sufficiently fast, (i.e. $a \sim e^{-(\rho z)^\alpha}$, with $\alpha >1$). A continuum without a mass gap is obtained for cases where $\hat{V} (z) \to 0$ for $z\to \infty$, like for AdS without an IR brane. Finally, the case of interest here is where a continuum appears with a mass gap $\mu$.  This corresponds to $\hat{V} (z) \to \mu$ for $z\to \infty$. The example corresponding to this case, which will be our canonical example for the AdS dual of the QCH, is
\begin{equation}
a(z) =\frac{R}{z} e^{-\frac{2}{3}\mu (z-R)} \, . 
\label{softwallmetric}
\end{equation}
It would be very interesting to understand what 4D CFT's and their necessary deformations are that correspond to such AdS duals, as well as how generic the case of a mass gap with a continuum is. These problems however are beyond the scope of this paper. 

In the following we will be focusing on the case of \eq{softwallmetric}.
By integrating over the bulk and using the solutions of the bulk equation of motion,  and rescaling by a factor 
\beq
\QH \equiv  L^{1/2} \,H_0~,\quad
\frac{1}{L} = Z_\Qh \frac{\Gamma(\Delta-1)}{\Gamma(2-\Delta)} R^{3-2\Delta} 2^{2\Delta-3}~, 
\eeq
to convert from 5D normalization to 4D normalization,
we obtain a 4D boundary effective theory for $\QH$.  
Generically with a bulk potential, $\QH$ has a VEV, which we will shift away as usual. In the unitary gauge we can write:
\beq
  \QH=\frac{1}{\sqrt{2}}\left(\begin{array}{c} 0 \\ {\mathcal V} + \Qh\end{array}\right) \, .
\eeq
With the appropriate background field, $\phi(z)$, we can reproduce \eq{quadraticaction} \cite{AdSCFTUnP,Falkowski}. For the soft-wall metric \eq{softwallmetric} this corresponds to 
\beq
\phi (z)= e^{\frac{4}{3}\mu (z-R)} \left(\frac{\nu^2-4}{R^2} -3\mu \frac{z}{R^2} \right) \, , 
\eeq
where $\nu = \sqrt{4 + m^2 R^2}$. In this case the normalized boundary-to-boundary propagator will be given by 
\begin{equation}
G_\Qh (R,R;p^2) = i \frac{1}{L} \left[ \frac{\mu K_{1-\nu}(\mu R)}{RK_\nu (\mu R)} - \frac{\sqrt{\mu^2-p^2}K_{1-\nu}(\sqrt{\mu^2-p^2}R)}{RK_\nu(\sqrt{\mu^2-p^2}R)} - M_0^2 \right]^{-1} \, , 
\label{adsbraneprop}
\end{equation}
where $K_\alpha$ is the modified Bessel function, and 
\beq
(M_0 R)^2 \approx \left\{ \left[(\mu^2-m_h^2) R^2\right]^{2-\Delta}- (\mu^2 R^2)^{2-\Delta} \right\} \, . 
\eeq
In the limit $p R, \mu R \ll 1$ \eq{adsbraneprop} reduces to \eq{quadraticaction}. The brane-to-bulk propagator, relevant for computing $n$-point correlators due to bulk interactions, is given by
\begin{equation}
G_\Qh (R,z;p^2) = G_\Qh (R,R;p^2) \cdot a^{-\frac{3}{2}}(z) (z/R)^\frac{1}{2} \frac{K_\nu(\sqrt{\mu^2-p^2}z)}{K_\nu(\sqrt{\mu^2-p^2}R)}\, .
\label{boundarytobulk}
\end{equation}
Different background fields will naturally yield different two-point correlators and different effective actions, corresponding to different models of IR breaking of conformality.  We choose this background as it results in an analytic two-point function, thus making the following discussion as transparent as possible.

 To obtain the appropriate Higgs VEV a bulk potential $V(H)$ must be included in ${\mathcal L}_\text{int}$, and other operators are allowed as well.  Once the two-point function is known, gauge invariance fixes the gauge interactions required by minimal coupling \cite{Unhiggs}, \ie the gauge interactions that saturate the Ward-Takahashi identities. \\

To obtain more general form factors we can include gauge invariant higher dimension operators in ${\mathcal L }_\text{int}$.
For example, if we include a higher dimension bulk operator that couples two gauge field strengths to the bulk Higgs,
\beq 
\frac{1}{M^2}\,H^\dagger  F_{\alpha\beta}F^{\alpha\beta} H \, , 
\eeq
 then we will have the corresponding 4D interaction in the 1PI effective action (a.k.a the boundary effective theory)
\beq
W \supset -(2\pi)^4\delta^4(p_1+p_2+p_3+p_4)g^2 c\,F_{VV\Qh\Qh}(p_i;\mu)F^a_{\alpha\beta}(p_1)F^{b\alpha\beta}(p_2)\QH^{\dagger}(p_3)T^a T^b \QH(p_4) ~.
\eeq
In the limit $p_i \gg \mu$, the form factor $F_{VV\Qh}(p_i;\mu)$ must become conformally invariant, and hence a falling function of momentum. The coupling should also vanish as $\mu \rightarrow 0$ if we want to recover a pure CFT. Setting one Higgs field to its VEV ($p=0$) yields an effective 4D vertex with two gauge bosons and one Higgs, that is a form factor  $F_{VV\Qh}(p_i;\mu)$ that can contribute to VBF, \eq{VBFformfactors}. In a soft wall AdS model with
a conformally flat metric,  taking flat zero-mode gauge bosons and with the boundary of AdS$_5$ at $z=R$, one finds that the effective 4D vertex is
\beq
 \delta^{ab} \left( g^{\alpha\beta} p_1\cdot p_2- p_1^{\beta} p_2^{\alpha}\right) F_{VV\Qh}~,
\label{VVhformfactor}
\eeq
where
\beq
F_{VV\Qh} =2 \frac{{\mathcal V} }{L\,M^2}  \, \int_R^\infty dz \,a^2 \left(\frac{z}{R}\right) \, \frac{K_{2-\Delta}(\sqrt{\mu^2-(p_1+p_2)^2} \, z) K_{2-\Delta}(  \mu \, z)}{K_{2-\Delta}(\sqrt{\mu^2-(p_1+p_2)^2} \, R) K_{2-\Delta}(  \mu \, R)} ~.
\label{hdformfactor}
\eeq
 This is obtained by propagating the Higgses from the boundary to the bulk using (\ref{boundarytobulk}) and inserting a VEV at zero momentum for one of them.

Another example of the type of form factor that can arise can be found in a generalized AdS model with
a bulk quartic interaction, 
\beq
\lambda_5 (H^\dagger   H)^2 \, ,
\eeq
which yields a quartic $\QH^4$ 4D coupling constant,
\beq
\lambda = \frac{ \lambda_5}{L^2}  \int_R^\infty dz \,\frac{1}{a} \left(\frac{z}{R}\right)^2 \, \left(\frac{K_{2-\Delta}(\sqrt{\mu^2-m_h^2} \, z) }{K_{2-\Delta}(\sqrt{\mu^2-m_h^2} \, R) }\right)^4 \, .
\label{quarticcoupling}
\eeq
In order to get the correct value of the the Higgs mass we must have (by equating the zero momentum limit of \eq{qptlag} to the negative of quadratic term of the shifted potential):
\beq
\lambda = \frac{\mu^{2 (2-\Delta)} -(\mu^2-m_h^2)^{2-\Delta} }{2{\mathcal V}^2 Z_\Qh} \, ,
\eeq
which reduces to the SM relation $\lambda = m_h^2/2 v^2$ in the limit $\Delta \rightarrow 1$, or in the limit $\mu\rightarrow \infty$.

\begin{figure}
\begin{center}
\includegraphics[width=.45\textwidth]{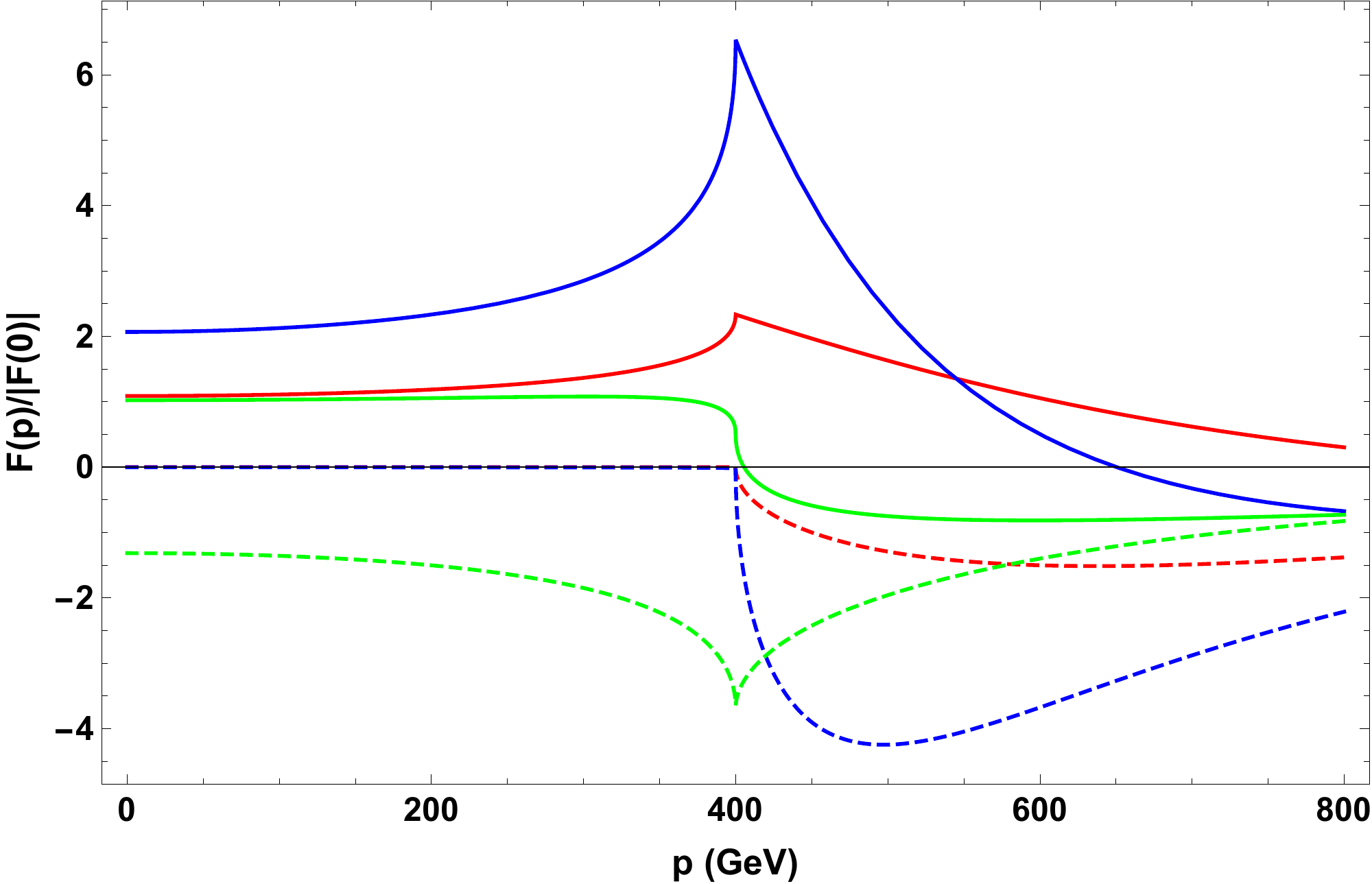}\hspace{0.5in}\includegraphics[width=.45\textwidth]{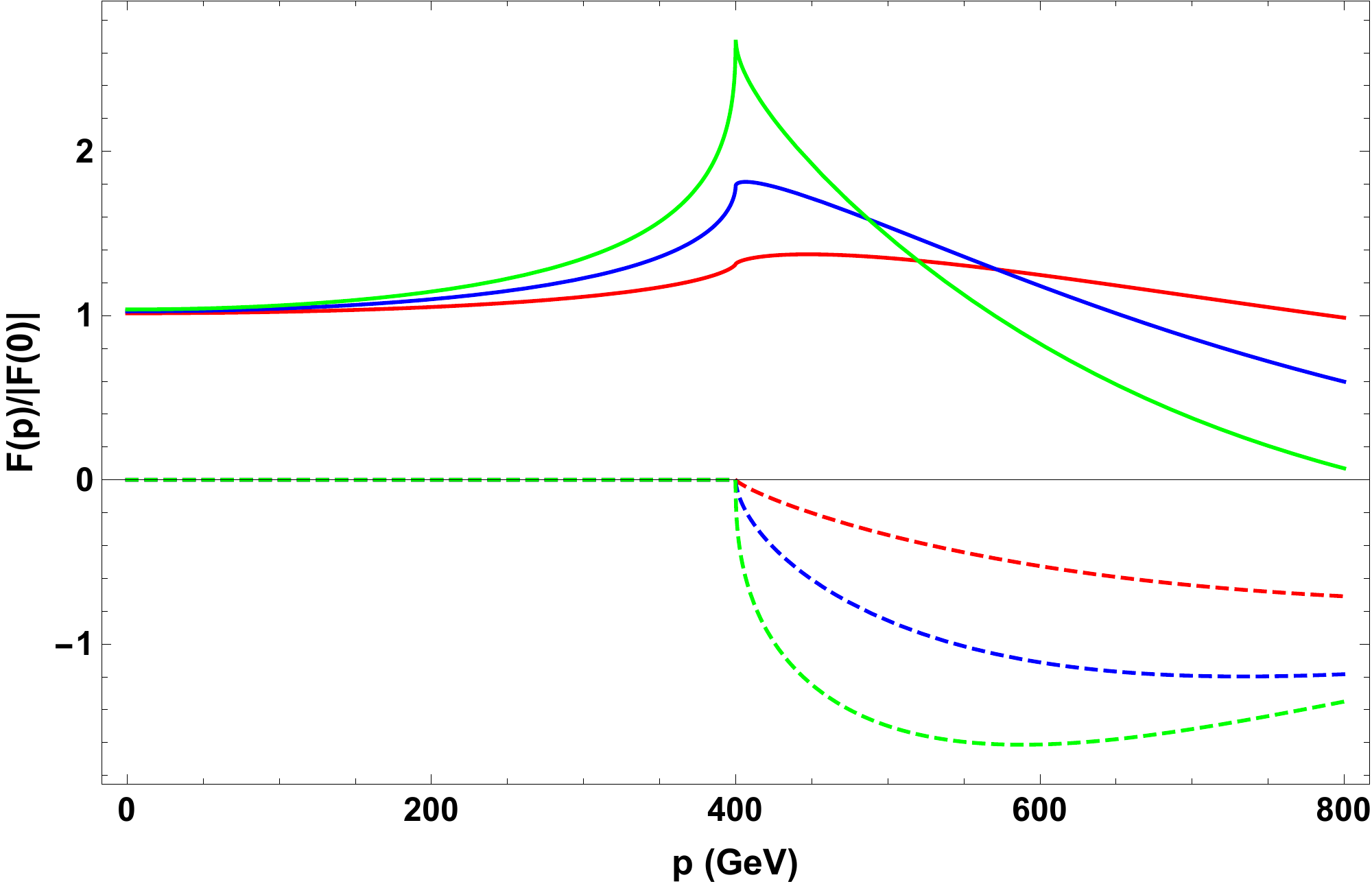}
\end{center}
\caption{Form factor for the cubic Higgs coupling, Eq.~(\ref{cubicformfactor}), for $\mu=400$, plotted  as a function of an off-shell Higgs momentum $p=\sqrt{p_\mu p^\mu}$.  The solid lines correspond to the real part of the form factor, and the dashed lines correspond to the imaginary part.  On the left, we keep only one Higgs on-shell.  We take $\Delta = 1.5$, and the three colors correspond to three different values of the off-shell Higgs momentum: $200$~GeV (red), $400$~GeV (blue), and $600$~GeV (green).  On the right, only one of the external Higgs fields are taken to be off-shell, and the colors correspond to $\Delta = 1.2$~(red), $1.4$~(blue), and $1.6$~(green).}
\label{fig:HHHformfactor}
\end{figure}

After setting one Higgs fields to its VEV this also yields a cubic 4D interaction, $\Qh^3$, with a form factor:
\beq
F_{\Qh\Qh\Qh} =\frac{ \lambda_5}{L^2}  \, {\mathcal V}    \int_R^\infty dz \,\frac{1}{a} \left(\frac{z}{R}\right)^2\, \frac{K_{2-\Delta}(\mu \, z)}{ K_{2-\Delta}(\mu \, R)}   \prod_{i=1}^3 \frac{K_{2-\Delta}(\sqrt{\mu^2-p_i^2} \, z) }{ K_{2-\Delta}(\sqrt{\mu^2-p_i^2} R)} \, .
\label{cubicformfactor}
\eeq
This is obtained by propagating the Higgses into the bulk using \eq{boundarytobulk} and inserting one VEV at zero momentum. An example is shown in Figure~\ref{fig:HHHformfactor}. 

Adding Yukawa interactions is also straightforward, as long as $\Delta <1.5$ \cite{Englert:2012cb}. In this case the Yukawa coupling is just the fermion mass divided by ${\mathcal V}$.

In both of these examples the form factor  at low momentum is almost constant and then peaks for momenta around $\mu$.  It would be very inefficient to describe the form factor by introducing higher dimension operators if the scale $\mu$ is within reach of the collider.

\section{Direct Signals of Quantum Criticality} \label{sec:collider}

A primary focus of Run II at the LHC will be to conduct detailed tests of Higgs phenomenology.  The signatures of quantum criticality can manifest in these analyses in several ways, including modifications of on-shell Higgs production and decay, drastic changes in the off-resonance high-momentum behavior of the Higgs two-point function due to e.g.~the continuum contributions, and finally modifications of $n$-point Higgs amplitudes which can result in sizable new physics contributions to, for example, double Higgs production. In the context of a non-mean-field theory description of electroweak symmetry breaking, collider studies of Higgs properties provide data on the scaling dimension of the operator that breaks electroweak symmetry, the threshold scale $\mu$, and on $n$-point CFT correlators.  

The production of new states above $\mu$ will modify the high-energy behavior $p^2 \gtrsim \mu^2$ of cross sections that involve the exchange of any of the Higgs components.  Both the neutral Higgs and the Goldstone bosons eaten by the $W$ and the $Z$ have propagators in the 1PI effective action that differ from the SM (see Eqs.~(\ref{propagator}) and (\ref{Gpropagator})).  In addition, the $n$-point correlation functions between neutral Higgs bosons and/or Goldstone bosons will have a form factor dependence that probes the manner in which the Higgs resonance arises, potentially distinguishing between models where the Higgs particle originates from or is mixed with the CFT.

An example of a potential signal for quantum criticality is the high energy behavior of the $gg \rightarrow ZZ$ process, which contributes to the ``golden" four-lepton signature.  At center of mass energies above the threshold for the cut in the Higgs two-point function, enhancements of the $gg \rightarrow h \rightarrow ZZ$ amplitude are expected.  As in the SM, the Higgs exchange diagrams interfere with a top-box diagram in which the $Z$ bosons are radiated off of virtual top quarks. This process has been studied extensively in the context of SM Higgs analyses~\cite{vanderBij:1988ac,Ellis:2014yca,Melnikov:2015laa}.  In Section~\ref{sec:offshell}, we describe an analysis of the differential rate of $gg \to ZZ$ that simultaneously probes the scaling dimension of $\Qh$ and the gap of the approximate CFT, $\mu$.

Another avenue to search for quantum criticality would be studies of the production of multiple on-shell Higgs bosons. The high luminosity LHC run is expected to begin probing double Higgs production towards the close of the LHC program, with a few events expected given SM calculations of the cross section.  If the Higgs originates as part of a CFT, or is perturbatively coupled to one, the continuum and/or the form factors associated with the CFT can give non-standard contributions to the double Higgs production amplitudes.  We will examine this possibility in Section~\ref{sec:double}.

\subsection{$ZZ$ production via a quantum critical Higgs} \label{sec:offshell}

The diagrams contributing to the $gg \to ZZ$ process are similar to the SM in the QCH framework. 
 There are two types of diagrams, one of which corresponds to a pure SM contribution \cite{vanderBij:1988ac}  without a Higgs exchange, and the usual gluon fusion diagram with an $s$-channel Higgs exchange:
\begin{center}
\includegraphics[width=4.5in]{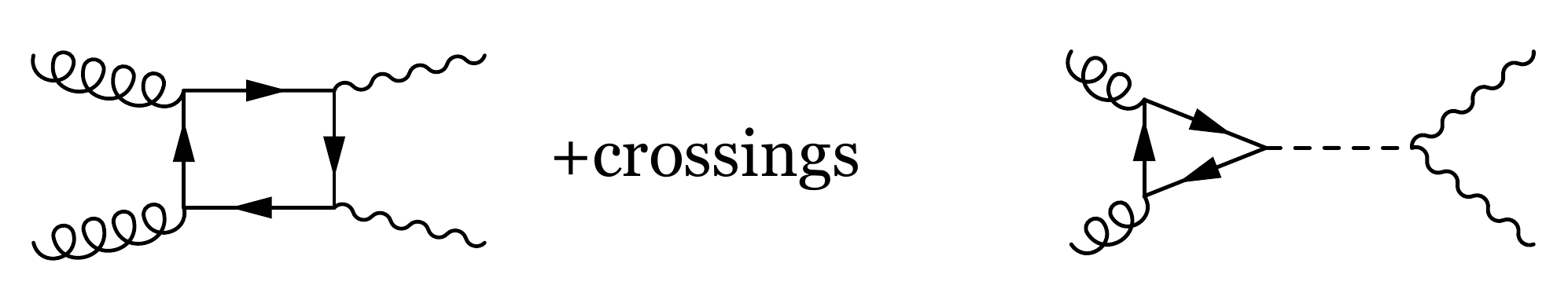}
\end{center}
The structures of the Higgs two-point function and of the $h ZZ$ form factor in the quantum critical case are modified, and as a consequence the interference between the two diagrams will be disturbed. We have discussed two scenarios in which we can obtain the form factors corresponding to dynamics in which there are new physics contributions to Higgs observables off the mass-shell.  We will focus on the case of minimal coupling, where a non-standard $hZZ$ form factor is present, \eq{hZZff}, being related by gauge invariance to the Higgs two-point function (see Appendix~\ref{app:A}).  Additional non-minimal form factors, like  the AdS bulk coupling form factor \eq{VVhformfactor}, could give additional contributions.

\begin{figure*}[!t]
\begin{center}
\includegraphics[width=.4\textwidth]{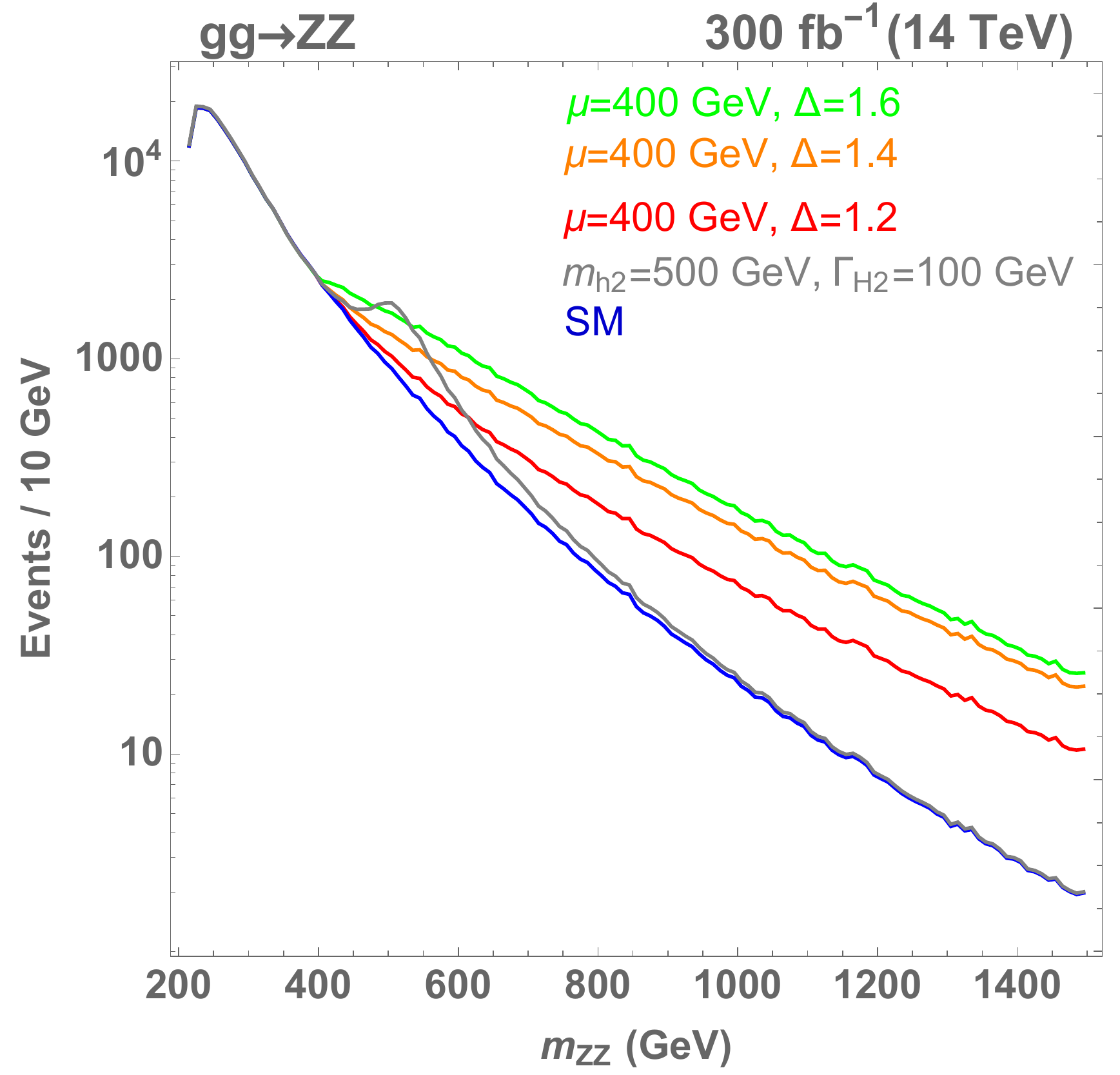}\;
\includegraphics[width=.4\textwidth]{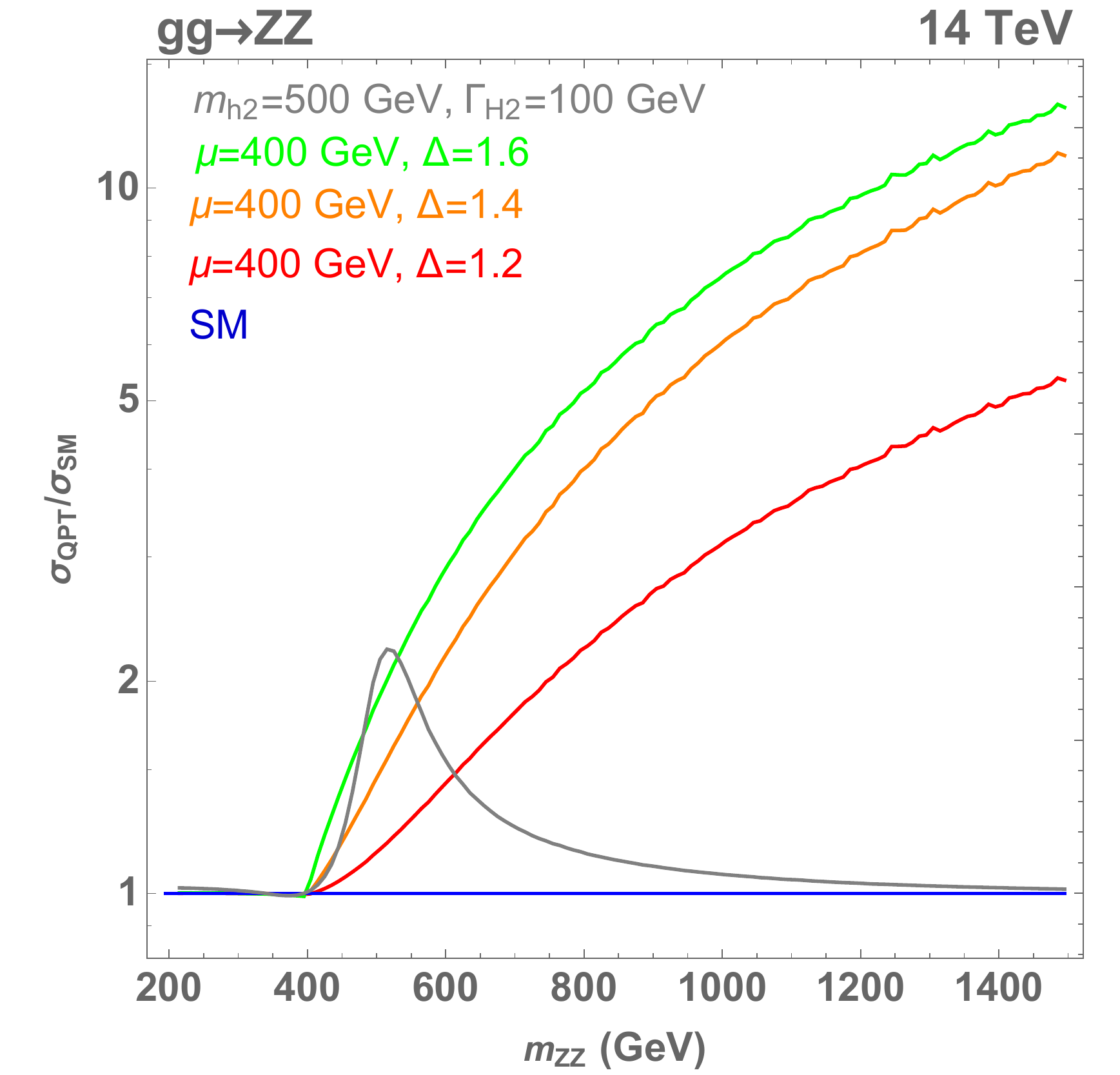}\;
\includegraphics[width=.4\textwidth]{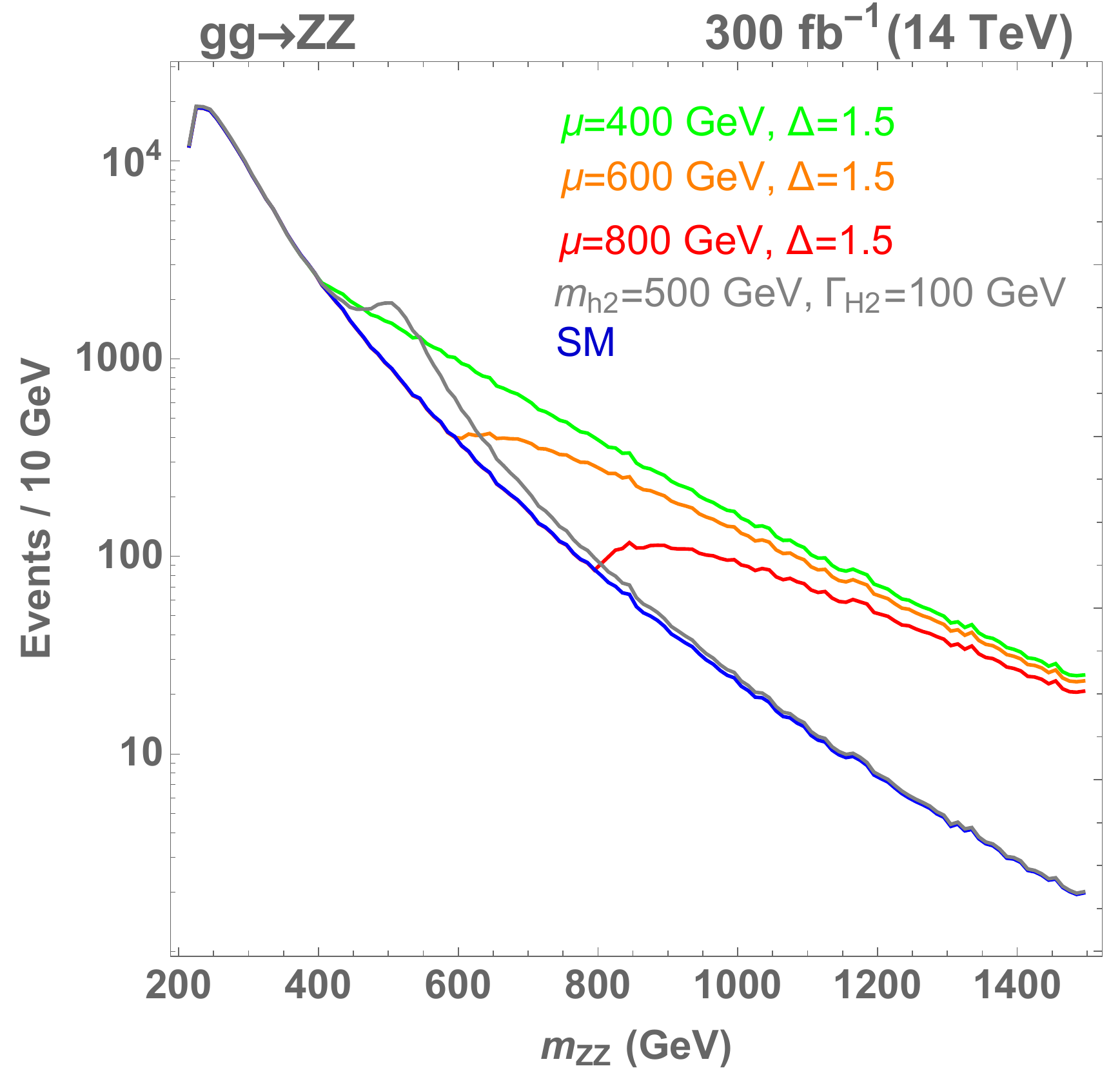}\;
\includegraphics[width=.4\textwidth]{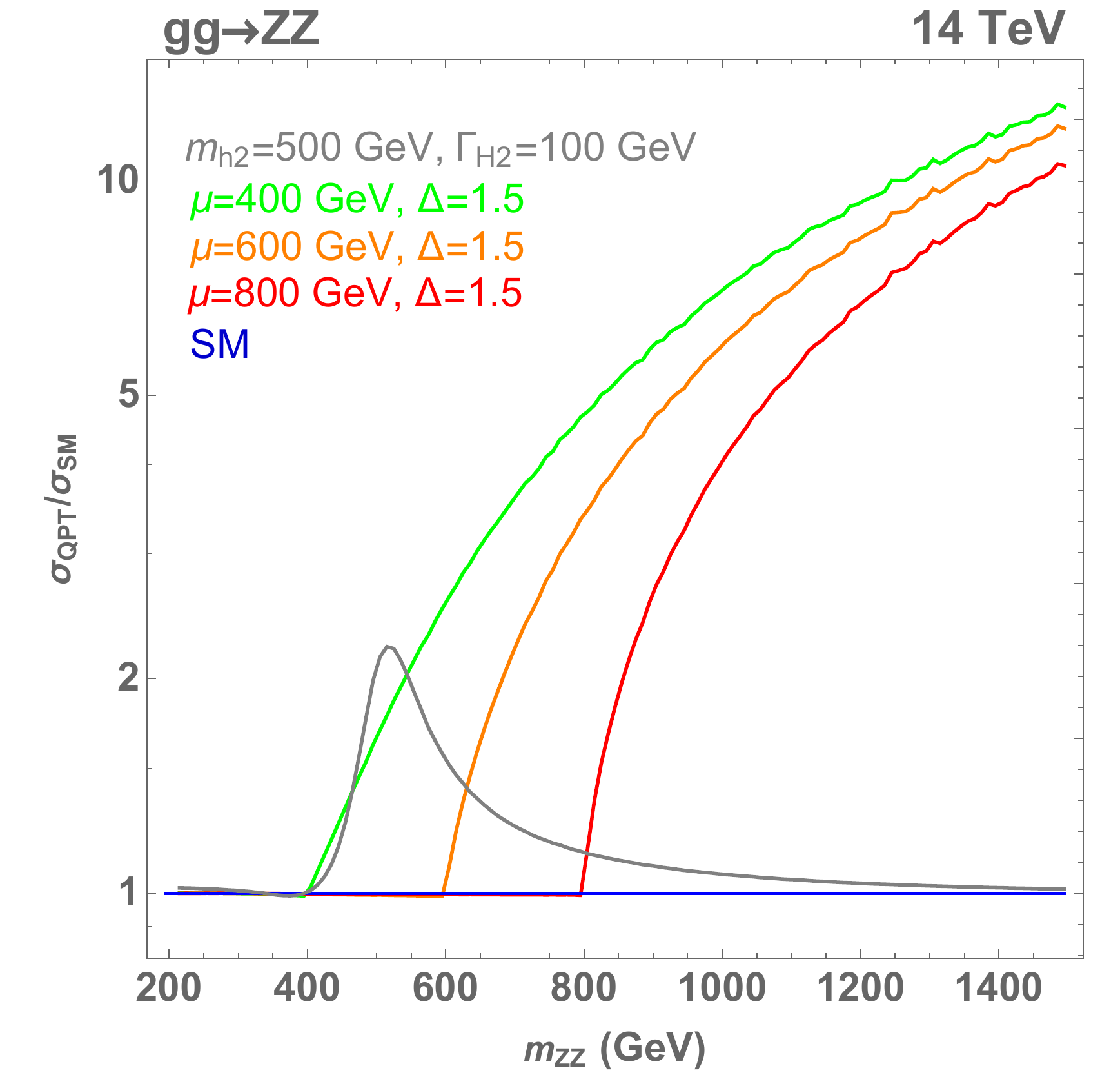}\;
\end{center}
\caption{The effects of including the QCH two-point function in the production of on-shell $Z$-boson pairs.  The effects of varying $\Delta$ (top two plots) and $\mu$ (bottom two plots) are shown.  On the right hand side, data are shown as a fraction of the SM result.  The effect of the cut associated with the non-standard two-point function is to dramatically enhance the production of $Z$ pairs far away from the Higgs pole. In each plot, for comparison purposes, the effect of a second heavy $500$~GeV Higgs with a $100$~GeV width is included.}
\label{fig:ZZ}
\end{figure*}

The expected number of $gg \rightarrow ZZ$ events (with on-shell $Z$-bosons) at $14$~TeV center of mass proton-proton collisions, with $300$~fb$^{-1}$ of integrated luminosity, are displayed in Figure~\ref{fig:ZZ}.  In these plots we vary both the scaling dimension of the QCH, $\Delta$, and the scale of conformal breaking, $\mu$.  Large enhancements over SM expectations are observed, particularly at large invariant masses of the $ZZ$ system, $m_{ZZ}$, where the contributions from the continuum are growing substantially.  For the phenomenologically viable choice of $\mu = 400$~GeV, an increase by a factor of 5-10 is typical at $m_{ZZ} \sim 1.5$~TeV.  We include, for comparison purposes, the corresponding distributions that would be generated if there were simply a second heavy neutral Higgs of mass $500$~GeV arising in a two Higgs doublet model with $\cos (\beta - \alpha) = 0.85$, maximizing the coupling of the heavy Higgs to $Z$ bosons while keeping the light Higgs couplings consistent with Run 1 LHC data. As expected the effect of the heavy Higgs quickly disappears from the invariant mass spectrum, while that of the continuum persists.

The reason for this large enhancement over the SM result is that the summation over diagrams in this process includes a cancellation of the leading behaviors of the box versus triangle amplitudes at large $s$.  The box and triangle diagrams are related by gauge invariance, and the cancellation occurs in order to maintain perturbative unitarity.  For this reason, the phenomenology is particularly sensitive to modifications of the Higgs two-point function, and the presence of the cut leads to a slower decrease of the amplitudes at higher $s$.

\subsection{Double quantum critical Higgs production} \label{sec:double}

The rate of Higgs pair production is an experimental probe that can potentially reveal the intrinsic nature of the Higgs. For example, in models~\cite{Witek,ourdilaton,Chacko,Kitano} where the Higgs arises from a conformal sector as the dilaton of spontaneously broken scale invariance, the Higgs cubic coupling could be $5/3$ that of the SM, even if all linear Higgs interactions are tuned to be precisely SM-like \cite{Witek}.  

Of course many new physics effects beyond the modification of the Higgs cubic coupling can play a role in the production of a pair of Higgs bosons at the LHC.  Other possibilities involve direct higher dimensional operators contributing to the effective $hhgg$ and $t \bar t hh$ vertices, modifications of the Higgs two-point function (as the ones we have been discussing), and also non-trivial $n$-point correlators due to the underlying strong dynamics.

\begin{figure*}[htb]
\begin{center}
\includegraphics[width=.45\textwidth]{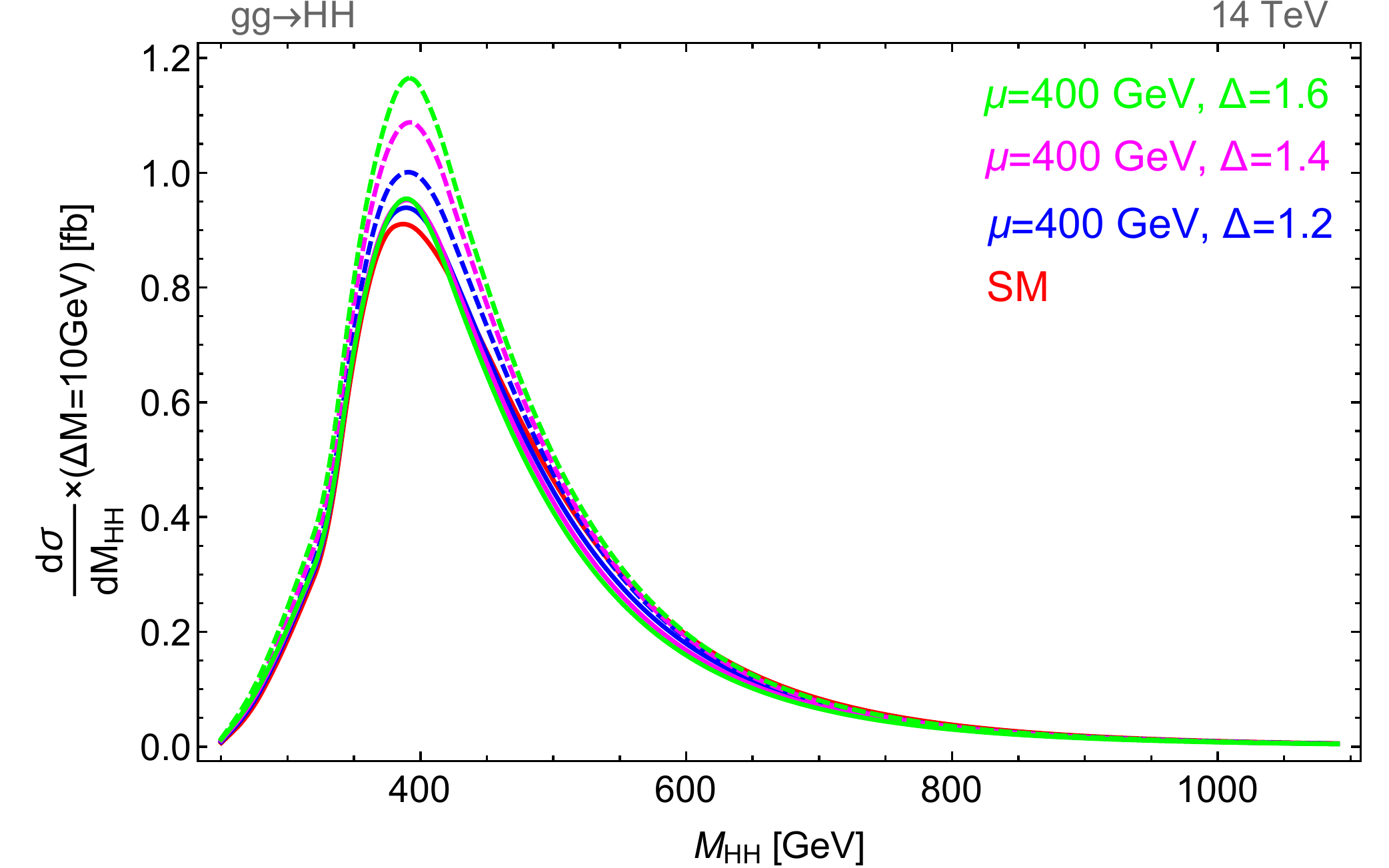}\;
\includegraphics[width=.45\textwidth]{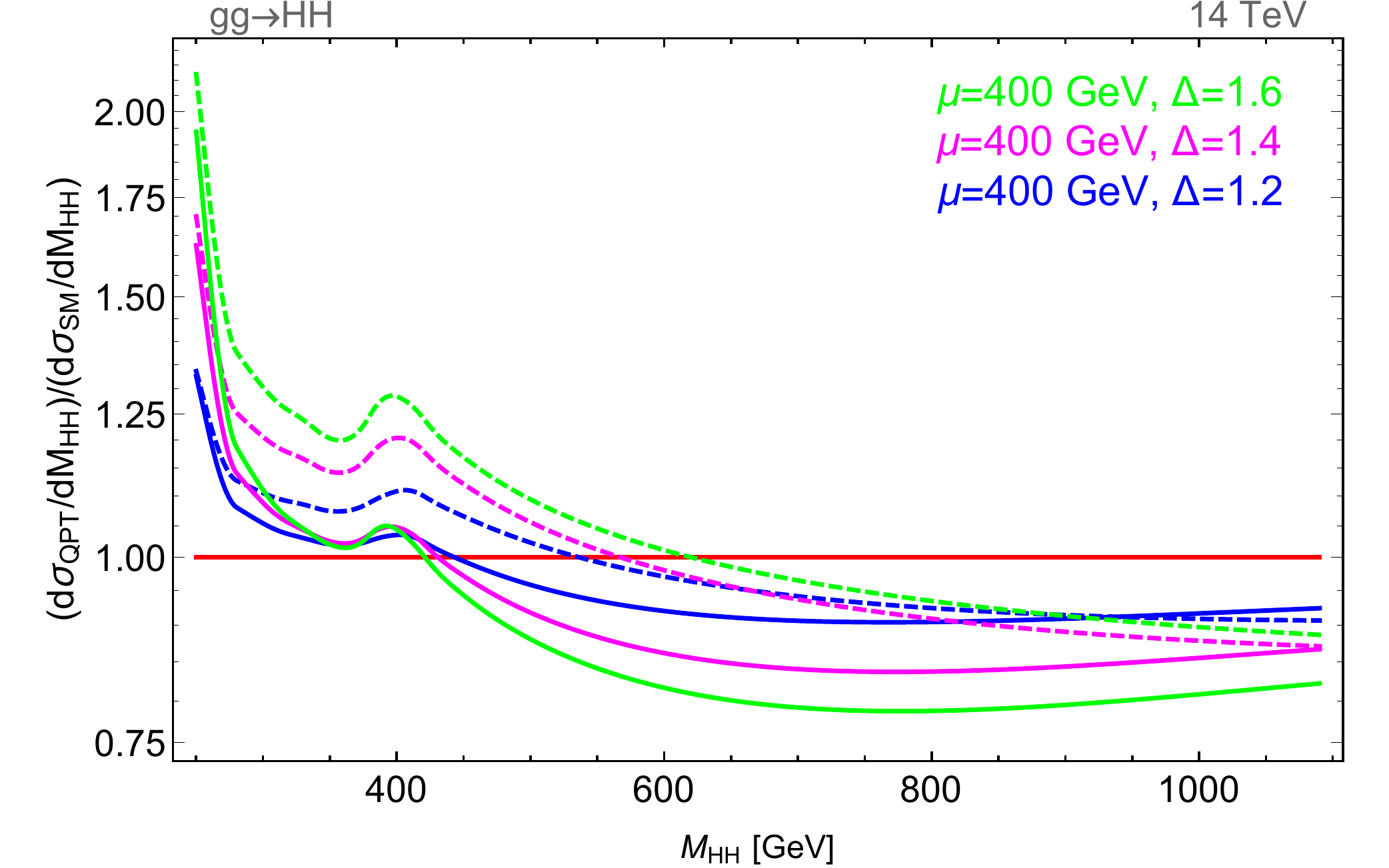}\;
\includegraphics[width=.45\textwidth]{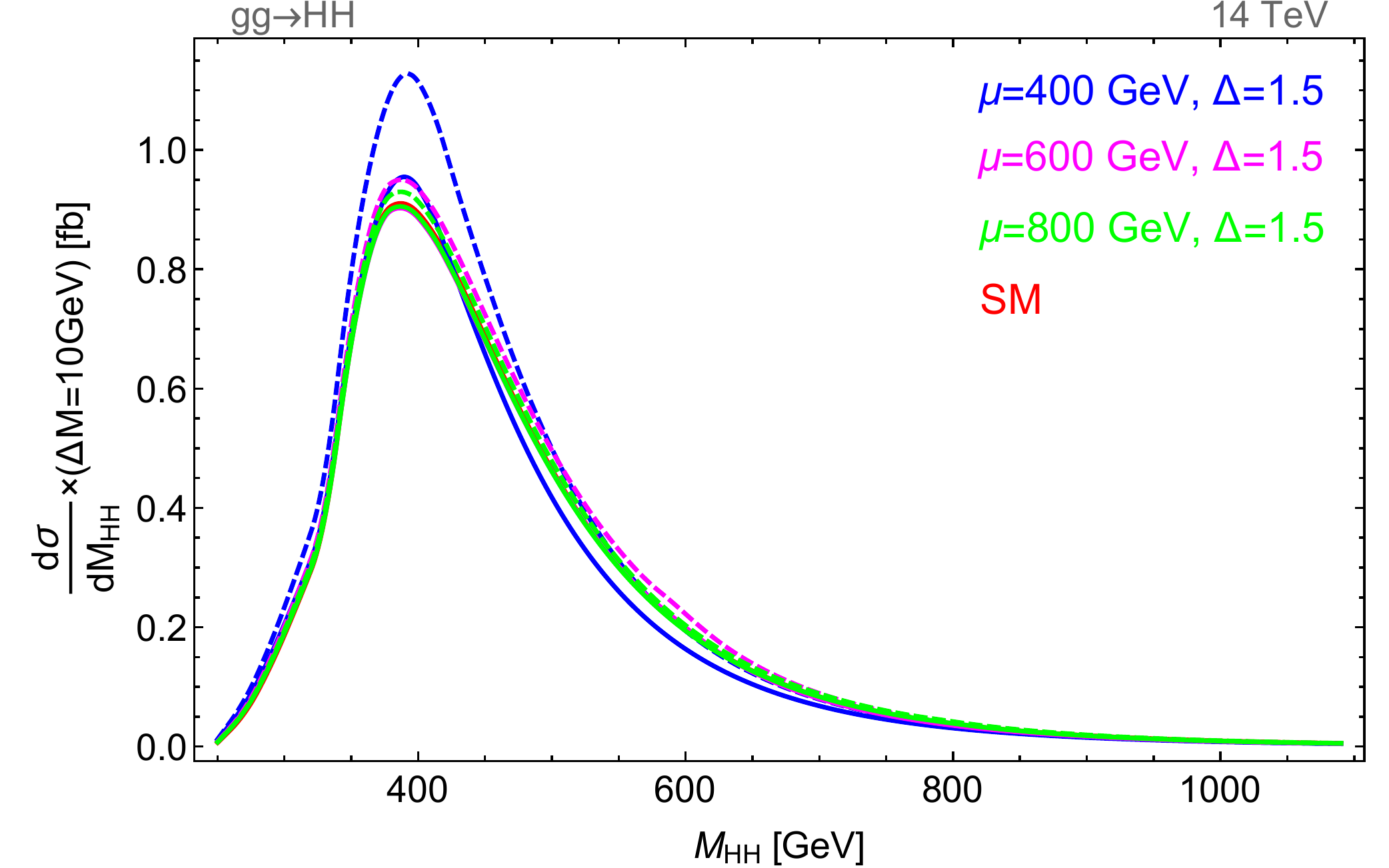}\;
\includegraphics[width=.45\textwidth]{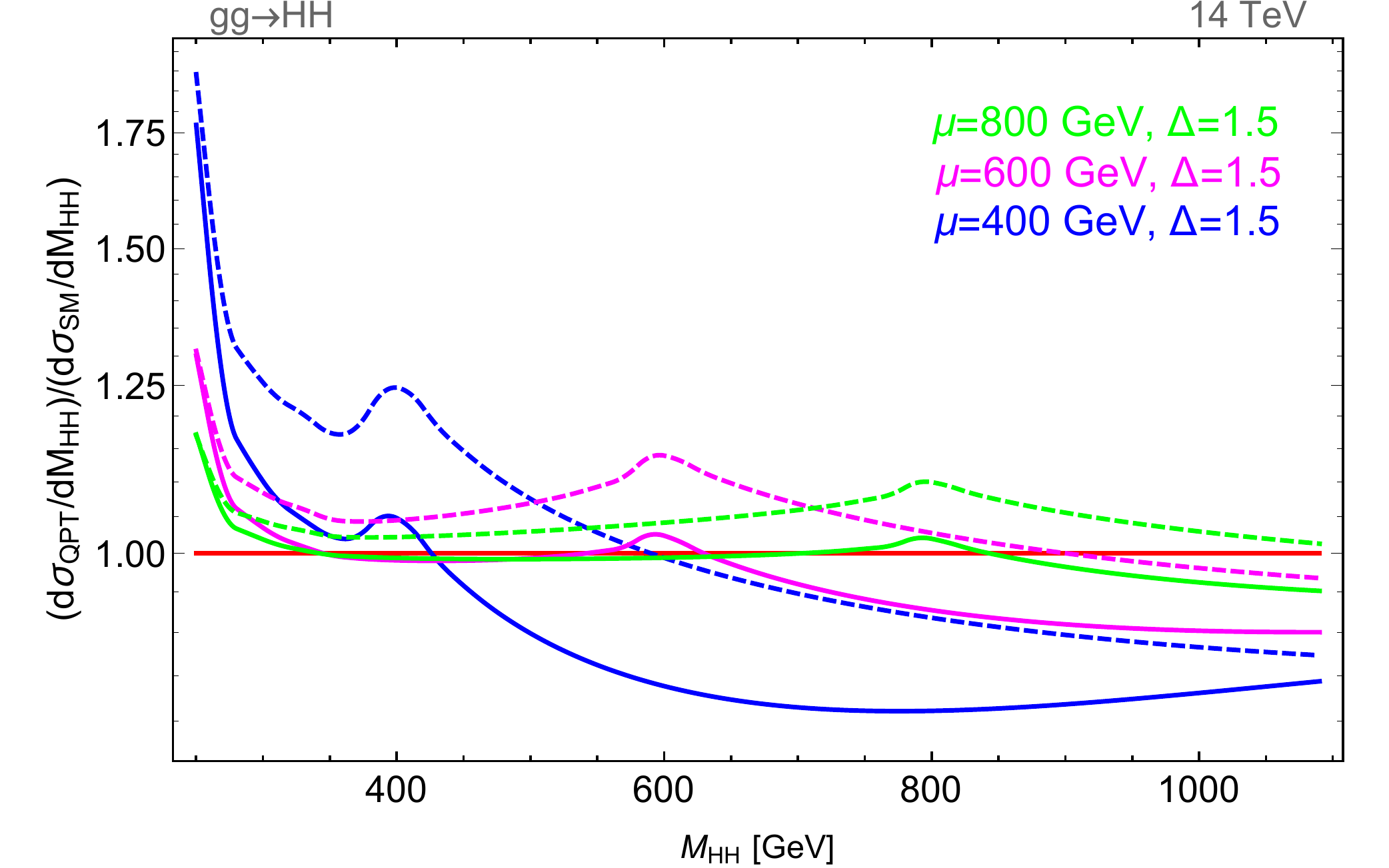}\;
\end{center}
\caption{Potential deviations in double Higgs production at the $14$~TeV LHC.  We consider two different cases.  The first (dashed lines) is when only the Higgs two-point function is modified, as when the strong sector that mixes with the Higgs has vanishing $n$-point correlators for $n\ge 2$.  The second (solid lines) is when the Higgs quartic coupling (and hence the cubic form factor) comes purely from a four-point correlator in a large $N$ CFT, and thus both the two- and three-point functions for the Higgs boson carry non-trivial momentum dependence. We vary $\Delta$ and $\mu$ for both cases, as shown. 
}
\label{fig:doubleHiggs}
\end{figure*}

In contrast to other processes, for the QCH, double production of on-shell Higgs bosons offers the opportunity to probe the higher $n$-point correlators of the CFT.  While it would be extremely interesting to see non-mean-field theory behavior in probes of the two-point function, the higher correlators encode information on the type of CFT we would be dealing with (e.g.~large $N$ theories, where the AdS/CFT correspondence offers a perturbative framework for estimating the higher-point correlators). 
In order to study such potential effects of quantum criticality, we computed the diagrams relevant for $hh$ production at the LHC.  These diagrams are similar to those associated with $ZZ$ production:
\beq
\includegraphics[width=4in]{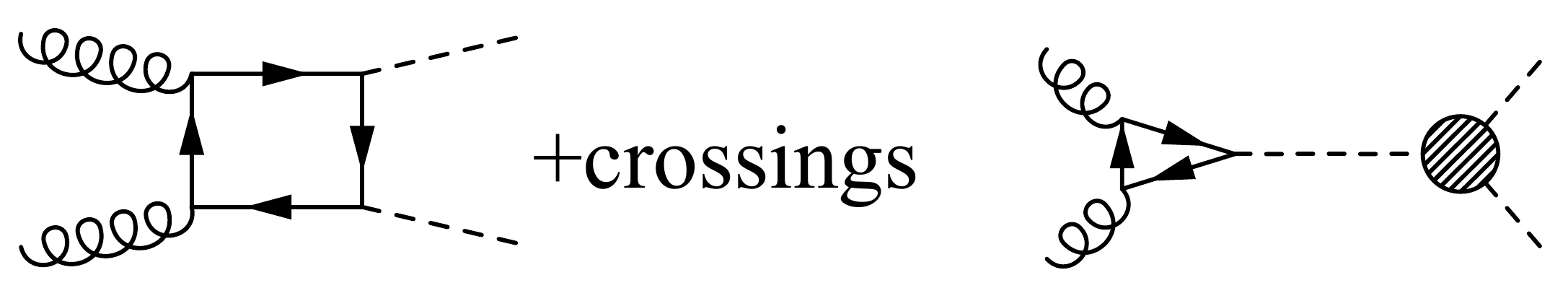}
\eeq
We examined both the case where there is a non-trivial form factor for the Higgs two-point and three-point functions, where the three-point function is shown in Eq.~(\ref{cubicformfactor}), and the case where the three-point function has no non-trivial momentum dependence. The results are displayed in Figure~\ref{fig:doubleHiggs}, where the solid lines correspond to the case with non-trivial three-point correlator. The dashed lines correspond to the case where only the Higgs two-point function shows a non-trivial behavior with momentum. 

Of particular note is the fact that the analysis of $Z$ pair production distributions performed in conjunction with studies of the double Higgs final state could help to  differentiate between the case of trivial and non-vanishing higher-$n$ correlation functions.  If electroweak symmetry breaking occurs via a QPT with non-mean-field behavior, some of the details of the CFT could be extracted from this data.

\section*{Conclusions} 

The puzzle of how the Higgs boson can be so light is still one of the greatest outstanding problems of particle physics, and recent LHC data have only made the problem more severe. In this paper we have taken a bottom-up approach: given that there is a light Higgs, what are the possible consistent low-energy theories? The SM is certainly the best known example; its crucial feature is that it can be tuned close to a quantum critical point. In general, being near a quantum critical point implies a hierarchy of scales, hence a long RG flow, and ultimately coming close to either a trivial fixed point (mean-field behavior) or a non-trivial fixed point (non-mean-field behavior). This suggests a large class of alternative possible theories: those with quantum critical points and non-mean-field behavior.  We have presented an effective theory that describes the low-energy physics of a broad class of such theories, with an arbitrary scaling dimension for the Higgs field. Gauge invariance requires that this scaling dimension also appears in form factors of the gauge couplings. We further showed how such effective theories can be constructed from an AdS$_5$ description, including the generation of form factors that are not determined by gauge invariance alone. Finally, we described how specific processes, $gg \rightarrow ZZ$ and double Higgs production can be used to gain information on the Higgs scaling dimension and form factor dependence, or put bounds on the mass threshold of the broken CFT states associated with the quantum critical point.

\section*{Acknowledgments}

We thank Zohar Komargodsky for useful discussions about generalized free field theories. We thank the Munich Institute for Astro- and Particle-Physics (MIAPP) of the DFG cluster of excellence ``Origin and Structure of the Universe'' for their hospitality and support while this work was in progress. C.C., B.B., and J.T.~thank the Aspen Center for Physics for its hospitality while this work was in progress.  B.B. thanks Mihail Mintchev for useful discussions about conformal field theory and generalized free fields. J.H.~and S.L.~thank the Cornell particle theory group for its hospitality during the completion of this this work. B.B.~is supported in part by the MIUR-FIRB grant RBFR12H1MW. C.C.~is supported in part by the NSF grant PHY-1316222.  J.H.~is supported in part by the DOE under grant DE-FG02-85ER40237. S.L.~is supported in part by Samsung Science \& Technology Foundation.
J.S.~has been supported in part by the ERC advanced grant 267985 (\emph{DaMeSyFla}). J.T.~is supported in part by the DOE under grant DE-SC-000999.

\appendix
\section{Form factors for generalized free fields: A minimal example}
\label{app:A}

The Lagrangian corresponding to the QPT presented in Section~\ref{sec:GFFT} can be written as
\beq
\Lag_{\QH} = - \frac{1}{Z_{\Qh}}\QH^\dagger \left[D^2+\mu^2 \right]^{2-\Delta}\QH - V(|\QH|) \, ,
\label{qptlagH}
\eeq
where $\QH$ is the QCH complex doublet. 
The Higgs potential, $V(|\QH|) + \mu^{4-2\Delta} |\QH|^2$, is such that the Higgs gets a VEV,
\beq
\vev{\QH}=\frac{1}{\sqrt{2}} \left(\begin{array}{c} 0 \\ {\mathcal V} \end{array}\right) \, ,
\eeq
breaking spontaneously the electroweak symmetry.
The Lagrangian for the excitation around the vacuum, $\Qh$, has been given in \eq{qptlag}.

From \eq{qptlagH} one finds for the trilinear interaction between two Higgses and one gauge boson, $\QH^{\dagger}(p+q) V_\mu^a(q) \QH(p) \Gamma^{\mu,a}(p,q)$, in momentum space:
\beq
\Gamma^{\nu, a}(p,q;\mu) = \frac{g}{Z_\Qh} T^a \left( 2 p^\nu+q^\nu \right) \Gamma(p, q;\mu) \, ,
\eeq
where the form factor reads
\beq
\Gamma(p, q;\mu) = - \frac{\left[ \mu^2 - (p+q)^2 \right]^{2-\Delta} - \left( \mu^2 - p^2 \right)^{2-\Delta}}{2 p \cdot q + q^2}  \, .
\eeq
One can explicitly check that when $\Delta = 1$ or $\mu \rightarrow \infty$, the above form factor reproduces the standard result $\Gamma^{\nu, a}_{\mathrm{SM}}(p,q) = g T^a (2 p^\nu+q^\nu$). 
The quartic interaction of two Higgses and two gauge bosons can be written as $\QH^{\dagger}(p+q+\bar q) V_\alpha^a(q) V_\beta^b(\bar q) \QH(p) \Gamma^{\alpha \beta , a b}(p,q,\bar q;\mu)$ in momentum space, where the form factor is given by
\beq
\Gamma^{\alpha \beta, a b}(p,q,\bar q;\mu) &=& 
\frac{g^2}{Z_\Qh} \bigg(  \{T^a,T^b\} g^{\alpha \beta} \Gamma(p, q + \bar q;\mu) \nonumber \\ 
&& + T^a T^b \frac{(2p+\bar q)^\beta (2 p + 2 \bar q + q)^\alpha}{q^2 + 2 (p + \bar q) \cdot q} \big[ \Gamma(p, q + \bar q;\mu) - \Gamma(p, \bar q;\mu) \big] \nonumber \\
&& + T^b T^a \frac{(2p+q)^\alpha (2 p + 2 q + \bar q)^\beta}{\bar q^2 + 2 (p + q) \cdot \bar q} \big[ \Gamma(p, q + \bar q;\mu) - \Gamma(p, q;\mu) \big] \bigg) \, .
\eeq
Again, for $\Delta = 1$ we obtain the SM result, $\Gamma^{\alpha \beta, a b}_{\mathrm{SM}}(p,q,\bar q) = g^2 \{T^a,T^b\} g^{\alpha \beta}$, and likewise in the limit $\mu^2 \gg p^2, q^2, \bar q ^2, p \cdot q, p \cdot \bar q, q \cdot \bar q$,
\bea
\Gamma^{\alpha \beta, a b}(p,q,\bar q;\mu) &\approx& 
g^2   \{T^a,T^b\} g^{\alpha \beta} \bigg(1 -\frac{m_h^2}{\mu^2} \bigg)^{\Delta -1} \, .
\eea
From this we can obtain the mass of the $W$,
\beq
m_W^2 = \frac{g^2  {\mathcal V}^2}{4} \left(1 -\frac{m_h^2}{\mu^2} \right)^{\Delta -1} \, .
\eeq
From the above interaction, the vertex of one Higgs boson $\Qh$ and two gauge bosons, $F_{VV\Qh}^{\alpha \beta, a b}(p,q,\bar q;\mu)$, is given by: 
\bea
F_{VV\Qh}^{\alpha \beta, a b}(p,q,\bar q;\mu) &=& 
\frac{g^2 {\mathcal V} }{Z_\Qh} \bigg(  \{T^a,T^b\} g^{\alpha \beta} \Gamma(0, q + \bar q;\mu) \nonumber \\ 
&& + T^a T^b \frac{\bar q^\beta (2 \bar q + q)^\alpha}{q^2 + 2 \bar q \cdot q} \big[ \Gamma(0, q + \bar q;\mu) - \Gamma(0, \bar q;\mu) \big] \nonumber \\
&& + T^b T^a \frac{q^\alpha ( 2 q + \bar q)^\beta}{\bar q^2 + 2 q\cdot \bar q} \big[ \Gamma(0, q + \bar q;\mu) - \Gamma(0, q;\mu) \big] \bigg) \nonumber \\
&& +\frac{g^2 {\mathcal V}}{Z_\Qh} \bigg(  \{T^a,T^b\} g^{\alpha \beta} \Gamma(-q - \bar q, q + \bar q;\mu) \nonumber \\ 
&& - T^a T^b \frac{(2q+\bar q)^\beta q^\alpha}{q^2} \big[ \Gamma(-q - \bar q, q + \bar q;\mu) - \Gamma(-q - \bar q, \bar q;\mu) \big] \nonumber \\
&& - T^b T^a \frac{(q+2\bar q)^\alpha \bar q^\beta}{\bar q^2} \big[ \Gamma(-q - \bar q, q + \bar q;\mu) - \Gamma(-q - \bar q, q;\mu) \big] \bigg) \, .
\label{hZZff}
\eea
The propagators for the $W$ and the $Z$ in unitary gauge are given by,
\beq
G_{W,Z}^{\alpha \beta} (p) = \frac{-i}{p^2-m_{W,Z}^2+i\epsilon} \left[  
g^{\alpha \beta} -\frac{p^\alpha p^\beta}{m_{W,Z}^2} \bigg( 1-\frac{p^2 - m_{W,Z}^2}{p^2} + \frac{(2-\Delta) (p^2-m_{W,Z}^2)}{\mu^2 \left[ 1 - \mu^{2 \Delta - 4} (\mu^2-p^2)^{2-\Delta} \right]} \bigg) \right] \, , 
 \label{Gpropagator}
\eeq
while that of the Higgs boson has been given in \eq{propagator}. In the limits $\mu^2 \gg p^2$ or $\Delta \rightarrow 1$ we recover the SM propagators in the unitary gauge.



\end{document}